\documentclass[prl,floatfix,showpacs,twocolumn,preprintnumbers,amsmath,amssymb,superscriptaddress]{revtex4}
\usepackage{siunitx}
\usepackage[sort&compress]{natbib}
\usepackage{graphicx,float,datetime}
\usepackage[ansinew]{inputenc}
\usepackage{hyperref, wasysym}
\usepackage{epstopdf, subfigure}
\usepackage{enumitem}
\usepackage{bm}

\newcommand*\xbar[1]{%
  \hbox{%
    \vbox{%
      \hrule height 0.5pt 
      \kern0.5ex
      \hbox{%
        \kern-0.1em
        \ensuremath{#1}%
        \kern-0.1em
      }%
    }%
  }%
} 

\newcommand{\udt}[3]{#1^{#2}_{\phantom{#2}#3}}

\newcommand{\dut}[3]{#1_{#2}^{\phantom{#2}#3}}

\newdateformat{mydate}{\THEDAY\hspace{3pt}\monthname[\THEMONTH] \THEYEAR}

\DeclareSIUnit\year{yr}

\allowdisplaybreaks[1]

\begin{document}

\begin{center}
\title{Stability of the flat FLRW metric in $f(T)$ gravity}
\date{\mydate\today}
\author{Gabriel Farrugia\footnote{gabriel.farrugia.11@um.edu.mt}}
\affiliation{Department of Physics, University of Malta, Msida, MSD 2080, Malta}
\affiliation{Institute of Space Sciences and Astronomy, University of Malta, Msida, MSD 2080, Malta}
\author{Jackson Levi Said\footnote{jackson.said@um.edu.mt}}
\affiliation{Department of Physics, University of Malta, Msida, MSD 2080, Malta}
\affiliation{Institute of Space Sciences and Astronomy, University of Malta, Msida, MSD 2080, Malta}

\begin{abstract}
{
In this paper, we investigate the stability of the flat FLRW metric in $f(T)$ gravity. This is achieved by analysing the small perturbations, $\delta$ about the Hubble parameter and the matter energy density, $\delta_\text{m}$. We find that $\delta \propto \dot{H}/H$ and $\delta_{\text{m}} \propto H$. Since the Hubble parameter depends on the function $f(T)$, two models were considered (A) the power-law model $f(T) = \alpha (-T)^n$, and (B) the exponential model $f(T) = \alpha T_0 \left(1 - \exp \left[-p \sqrt{\dfrac{T}{T_0}}\right]\right)$, where the parameters $n$ and $p$ were chosen to give comparable physical results. For the parameters considered, it was found that the solutions are stable with vanishing $\delta$ and decaying then constant $\delta_{\text{m}}$, meaning that the matter perturbations persist during late times.
}
\end{abstract}

\pacs{}

\maketitle

\end{center}

\section{I. Introduction}\label{sec:intro}

Amendments to Einstein's general theory of relativity (GR) have been under way almost from the inception of the model of gravity. With the discovery of the accelerating universe, the importance of dark energy and the cosmological constant as means to describe this phenomenon became one of the biggest unsolved problems in cosmology \cite{Riess:1998cb,Perlmutter:1998np,Hinshaw:2012aka,deSabbata:1990rn,Peebles:2002gy}. Furthermore, the introduction of the unknown mass density known as dark matter as a way to correct the rotation curves of galaxies is also a major problem. Alternative proposals without invoking unknown matter fields has been the context of modified and alternative theories of gravity. One example of this is the instance of $f(R)$ gravity where the Ricci scalar, $R$, in the action Lagrangian is replaced by an arbitrary function of $R$ (a detailed review on $f(R)$ gravity is given in Ref. \cite{DeFelice:2010aj} and references therein). 

GR describes gravity through the concept of spacetime curvature which is how gravity exhibits itself. Besides the curvature notion of gravity, there has been a lot of work involved in another formulation of gravity called teleparallel gravity \cite{DeAndrade:2000sf,aldrovandi2012teleparallel}. In this reformulated theory of gravity, curvature no longer describes gravity and is replaced by torsional quantities. This teleparallel formulation makes use of a different theoretical foundation, with the Ricci scalar being replaced by a torsion scalar $T$ (not to be confused with the trace of the stress-energy tensor, $\mathcal{T}$), but the theory turns out to be equivalent to that of GR [called Teleparallel Equivalent of General Relativity (TEGR)], up to a boundary term difference \cite{Garecki:2010jj,Maluf:2013gaa,Arcos:2005ec}.

Although the theory is still general covariant, the model has some noticeable differences from that of GR. One such differences arises from the independent degrees of freedom; although the metric tensor has 10 independent degrees of freedom as in GR, the vierbeins, which construct the latter, have 16 independent degrees of freedom. These 6 extra degrees of freedom were found to be related to the inertial effects of the system. This in turn seemingly resulted in having the theory no longer local Lorentz invariant but it was then realised that the field equations were local Lorentz invariant \cite{Obukhov:2006sk}.

This alternative formulation of GR resulted in a modified theory, called $f(T)$ gravity, where the torsion scalar in the action is generalised to a general function of it. In this way, this theory becomes analogous to that of $f(R)$ gravity, with the advantage that the theory was no longer fourth order but second order. However, the theory seemed to suffer from lack of local Lorentz invariance even in the field equations (see \cite{Cai:2015emx} and references therein). This resulted into a large investigation and a notion of choosing the right observer when studying the theory (the idea of what are called good tetrads and bad tetrads, see Ref. \cite{Tamanini:2012hg}). 

Recently however, it was discovered that $f(T)$ gravity can in fact be a general covariant theory (i.e. satisfied local Lorentz invariance). As Kr\v{s}\v{s}\'{a}k and Saridakis argue in Ref. \cite{Krssak:2015oua}, the problem in the original formulation where it was assumed that the spin connection (which contains information about inertial effects) to be vanishing in all frames. However, some tetrads (`bad' tetrads) did not have a vanishing spin connection resulting in the wrong set of field equations. In the work, the authors devise a method to be able to determine such spin connection and make the theory covariant. The good tetrads then become special cases of the theory, where such tetrads would give a vanishing spin connection.

Motivated by this, the theory of teleparallelism could in turn be a viable alternative theory of gravity. The study of this theory in cosmology has been studied in some recent works (see Ref. \cite{Cai:2015emx} and references therein). In this paper, we investigate one aspect of the theory, that of its stability in an expanding homogeneous and isotropic universe. The study of stability helps to constrain the possible allowed function of the theory (for example in $f(R,G)$ Ref. \cite{delaCruzDombriz:2011wn} and $f(R,\mathcal{T})$ theory \cite{Sharif:2014ioa} respectively). Stability has also been studied in teleparallelism, within the topics of reconstruction, thermodynamics and stability [$f(T)$ gravity in Ref. \cite{Salako:2013gka} and $f(T,\mathcal{T})$ gravity in Ref. \cite{Junior:2015bva}]. Furthermore, stability in scalar perturbations and coupled scalar fields can be found in Refs. \cite{Harko:2014aja,Biswas:2015cva}. 

In this paper, we are interested in the stability of the background evolution of the universe in a homogeneous and isotropic universe in $f(T)$ gravity. In particular, we derive the analytical solution for the evolution of the perturbation variables for general $f(T)$ functions, in contrast to what is found in Ref. \cite{Salako:2013gka} where the evolution function obtained are for the reconstruction models considered. Furthermore, various other models are able to mimic $\Lambda$CDM evolution (can be found in Ref. \cite{Nesseris:2013jea}) but their stability analysis has not been carried out. As such, we investigate two such $f(T)$ models to study their stability.

The paper is divided as follows, a brief overview of $f(T)$ gravity is given in Sec. II, followed by the derivation of the perturbed Hubble parameter and energy density in Sec. III. Afterwards, the power-law and de-Sitter scale factor evolutions are analysed in Sec. IV. In Sec. V, the two models considered in this paper are analysed followed finally by a conclusion in Sec. VI.

\section{II. An overview of $f(T)$ gravity}\label{sec:field-equations}

\subsection{A. Connections, action and field equations}

The theory of teleparallelism requires a new starting point from that of GR. Curvature in GR is obtained through the use of the Levi-Civita connection (which is torsion-free), and hence a new connection is needed for teleparallel gravity, one which is curvature-free. This is the Weitzenb\"{o}ck connection $\widehat{\Gamma}^{\alpha}_{\mu\nu}$, which is defined as
\begin{equation}\label{eq:weitzenbockdef}
\widehat{\Gamma}^{\rho}_{\nu\mu} \equiv \dut{e}{a}{\rho}\partial_\mu \udt{e}{a}{\nu} + \dut{e}{a}{\rho}\udt{\omega}{a}{b\mu}\udt{e}{b}{\nu},
\end{equation}
where ${e^a}_\rho$ and ${e_a}^\mu$ are referred to as vierbeins (or tetrads) along with their respective inverses, and $\udt{\omega}{a}{b\mu}$ is called the \textit{purely inertial spin connection} which is related to the inertial effects of the system under consideration \cite{aldrovandi2012teleparallel,Krssak:2015lba}. The two indices refer to two coordinate systems; the Latin indices transform like an inertial spacetime coordinate, while the Greek indices transform like global coordinates. In this way, these vierbeins can be used to relate to the metric tensor $g_{\mu\nu}$ depending on the local position $x$ on the spacetime manifold through
\begin{equation}
g_{\mu\nu}\left(x\right)\equiv \udt{e}{a}{\mu}\left(x\right)\udt{e}{b}{\nu}\left(x\right)\eta_{ab},
\end{equation}
where $\eta_{ab}$ is the Minkowski metric tensor diag$(1,-1,-1,-1)$. Thus, the vierbein links the local Minkowski metric to the global metric tensor. From this point onward, the explicit expression of a local position $x$ will be suppressed for brevity's sake. 

The Riemann tensor (which quantifies curvature) is replaced with the \textit{torsion tensor} (which quantifies torsion) and is defined to by
\begin{equation}\label{eq:torsiontensordef}
\udt{T}{a}{\mu\nu} \equiv \partial_\mu \udt{e}{a}{\nu} - \partial_\nu \udt{e}{a}{\mu} + \udt{\omega}{a}{b\mu}\udt{e}{b}{\nu} - \udt{\omega}{a}{b\nu}\udt{e}{b}{\mu}.
\end{equation}
Using the torsion tensor, the \textit{superpotential tensor} is defined by
\begin{equation}\label{eq:superpotentialdef}
\dut{S}{a}{\mu\nu}\equiv \frac{1}{2}\left(\udt{K}{\mu\nu}{a}+\dut{e}{a}{\mu}\udt{T}{\alpha\nu}{\alpha}-\dut{e}{a}{\nu}\udt{T}{\alpha\mu}{\alpha}\right),
\end{equation}
where $\udt{K}{\mu\nu}{a}$ is the \textit{contorsion tensor} defined as
\begin{equation}\label{eq:contorsiondef}
\udt{K}{\mu\nu}{a} \equiv \dfrac{1}{2} \left(\dut{T}{a}{\mu\nu} + \udt{T}{\nu\mu}{a} - \udt{T}{\mu\nu}{a}\right).
\end{equation}
Using Eq. \eqref{eq:torsiontensordef} and \eqref{eq:superpotentialdef} leads to the torsion scalar through
\begin{equation}\label{eq:torsionscalardef}
T \equiv \dut{S}{a}{\mu\nu}\udt{T}{a}{\mu\nu},
\end{equation}
which defines the action for teleparallel gravity to be
\begin{equation}\label{eq:teleparallel-action}
S = \dfrac{1}{16\pi G} \int d^4x \: e \: T + \int d^4x \: e \: \mathcal{L}_m,
\end{equation}
where $e = \det\left(\dut{e}{\mu}{A}\right) = \sqrt{-g}$ and $\mathcal{L}_m$ is the matter Lagrangian. This becomes the action equivalent to GR (i.e. TEGR) which is general covariant. In the same way as the Ricci scalar in GR is generalised to some general function $f(R)$, the TEGR action is generalised to a general torsion function $f(T)$
\begin{equation} \label{eq:general-action}
S = \dfrac{1}{16\pi G} \int d^4x \: e \: \left[T + f(T)\right] + \int d^4x \: e \: \mathcal{L}_m.
\end{equation}
By varying the action with respect to the vierbein field, the following field equations are obtained \cite{Krssak:2015oua}
\begin{widetext}

\begin{equation}\label{eq:general-field-equations}
\left(1+f_T\right) \left[e^{-1} \partial_\nu \left(e \dut{S}{a}{\mu\nu}\right) - \udt{T}{b}{\nu a} \dut{S}{b}{\nu\mu} + \udt{\omega}{b}{a\nu}\dut{S}{b}{\nu\mu} \right]+ f_{TT} \dut{S}{a}{\mu\nu} \partial_\nu T + \udt{e}{\rho}{a} \left(\dfrac{f+T}{4}\right) = 4\pi G \udt{e}{\alpha}{a} \stackrel{\textbf{em}}{\dut{T}{\alpha}{\rho}}.
\end{equation}

\end{widetext}
where $\stackrel{\textbf{em}}{\dut{T}{\alpha}{\rho}}$ is the stress-energy tensor, which in terms of the matter Lagrangian is given by $\stackrel{\textbf{em}}{\dut{T}{\beta}{\rho}} = \dfrac{1}{e} \dut{e}{\beta}{a} \dfrac{\delta \left(e \mathcal{L}_{\text{m}}\right)}{\delta \udt{e}{a}{\rho}}$ with $\delta$ representing the vierbein perturbation.

These field equations are general covariant and hence frame independent, removing the local Lorentz invariance issue originally present in the theory. In the proper tetrad formalism, the spin connection becomes zero and reduces to the field equations found in Refs. \cite{Harko:2014aja,Saez-Gomez:2016wxb}, where the pure tetrad formalism is used to derive the equations (where the spin connection is assumed to be zero \textit{a priori}). The proper tetrad formalism is still general covariant since it does not assume the spin connection to be zero \textit{a priori}, and is just a special case of the covariant formulation of $f(T)$ gravity. As such, in the work which follows, the proper tetrad formalism is used and the spin connection is allowed to vanish.

\subsection{B. Flat, isotropic and homogeneous universe in $f(T)$ gravity}

As described in Sec. I, observational data suggests that the universe is generally flat, isotropic and homogeneous. Hence, we are only interested in one particular metric, that being the spatially flat Friedmann-Lemaitre-Robertson-Walker (FLRW) metric
\begin{equation}\label{eq:flat-FLRW}
ds^2 = dt^2 - a^2(t)\left(dx^2+dy^2+dz^2\right),
\end{equation}
where $a(t)$ is the scale factor in terms of cosmic time. For such a metric, a diagonal vierbein field of the form
\begin{equation}\label{eq:diag-tetrad}
\dut{e}{\mu}{a} = \text{diag}\left(1,a(t),a(t),a(t)\right),
\end{equation}
is considered, which is a pure tetrad. In this case, $T = -6H^2$. Using the field equations in Eq. \eqref{eq:general-field-equations}, this gives rise to the two GR modified equations
\begin{align}
& f-T-2Tf_T = 2 \kappa^2 \rho, \label{eq:00-zero} \\ 
& \dot{H} = -\dfrac{\kappa^2 \left(\rho+p\right)}{2\left(1+f_T + 2T f_{TT}\right)}, \label{eq:trace-zero}
\end{align}
where the $tt$-equation was used to simplify the spatial equation. Note that the spatial equation holds only if $1+f_T + 2T f_{TT} \neq 0$. In other words, this is valid provided $f(T) \neq -T + c_1 \sqrt{-T} + c_2$, where $c_1$ and $c_2$ are integration constants. The square root term does not play a role in the equations and hence is neglected [this term gives the same expansion history as Dvali, Gabadadze and Porrati (DGP) gravity (see their paper Ref. \cite{Dvali:2000hr} for more details) and thus only plays a role in higher dimensional theories \cite{Linder:2010py}). The remaining terms reduces the action to $c_2$, which plays the role of a cosmological constant. This leads to a non-physical evolution and hence this case is neglected. 

The modified Friedmann equations can be used to derive the continuity equation
\begin{equation}\label{eq:continuity-zero}
\dot{\rho} + 3H\left(\rho+p\right) = 0,
\end{equation}
which is essentially the same as GR. By defining an equation of state (EoS) parameter $w$ for the matter component through the relation
\begin{equation}\label{eq:eos-matter}
p = w\rho,
\end{equation}
the continuity equation can be solved to give
\begin{equation}\label{eq:continuity-zero-solution}
\rho = \rho_0 {a(t)}^{-3(1+w)},
\end{equation}
provided that $w$ is constant in time.

By examining the modified Friedmann equations, one notes that the extra components arising from $f(T)$ gravity can be used to define an exotic fluid having energy density $\rho_{\text{exo}}$ and pressure $p_{\text{exo}}$
\begin{align}
&\kappa^2 \rho_{\text{exo}} \equiv Tf_T - \dfrac{f}{2}, \label{eq:def-DE-energy-density} \\
&\kappa^2 p_{\text{exo}} \equiv -\kappa^2 \rho_{\text{exo}} + 2\dot{H}\left(f_T+2T f_{TT}\right). \label{eq:def-DE-pressure}
\end{align}
which in turn can be used to define an effective equation of state parameter $w_{\text{exo}}$ to be
\begin{align}
&w_{\text{exo}} \equiv \dfrac{p_{\text{exo}}}{\rho_{\text{exo}}} = -1 - 4\dot{H}\dfrac{f_T+2T f_{TT}}{f-2Tf_T}.
\end{align}
This can expressed in terms of $f(T)$ only by using the field equations Eqs. \eqref{eq:00-zero} and \eqref{eq:trace-zero}, and then in terms of the EoS parameter for matter Eq. \eqref{eq:eos-matter} to give
\resizebox{0.95\linewidth}{!}{
  \begin{minipage}{\linewidth}
\begin{equation}\label{eq:def-EoS-DE}
w_{\text{exo}}= -1 + (1+w) \dfrac{\left(f-T-2T f_T\right)\left(f_T+2T f_{TT}\right)}{\left(1 + f_T + 2T f_{TT}\right)\left(f-2Tf_T\right)}.
\end{equation}
  \end{minipage}
}
In this way, the Friedmann equations can be reduced into a more familiar form, 
\begin{align}
-T &= 2\kappa^2 \left(\rho + \rho_{\text{exo}}\right), \\ 
2\dot{H} &= -\kappa^2 \left(\rho+p+\rho_{\text{exo}} + p_{\text{exo}}\right).
\end{align}
Using this reformulation of the field equations, a continuity equation for the exotic fluid similar to the matter content can also be derived, giving
\begin{equation}
\dot{\rho}_{\text{exo}} + 3H\left(\rho_{\text{exo}} + p_{\text{exo}}\right) = 0.
\end{equation}

Another key component in describing the evolution of the universe is the deceleration parameter which is defined to by
\begin{equation}\label{eq:def-deceleration-parameter}
q(t) \equiv -\dfrac{\ddot{a}a}{{\dot{a}}^2} = -\dfrac{\dot{H}}{H^2} -1,
\end{equation}
which for $f(T)$ gravity is expressed as
\begin{equation}
q(t) = -1+\frac{3 (1+w) \left(T+2 T f_T-f\right)}{2 T \left(1+f_T+2 T f_{TT}\right)}.
\end{equation}
Evidence shows that the universe is expanding, it implies that $q\left(t_0\right) < 0$. This criterion is important when analysing the models in Secs. V and VI.

\section{III. Perturbations of the flat FLRW metric in $f(T)$ gravity}

In this section, we shall consider perturbations of the homogeneous and isotropic FLRW metric and study their evolution, which ultimately determines whether the cosmological solutions in $f(T)$ gravity are stable. The perturbations considered are of first order, and are described by
\begin{equation}
H(t) \rightarrow H(t)\left(1+\delta\right), \: \rho(t) \rightarrow \rho(t)\left(1+\delta_{\text{m}}\right),
\end{equation}
where $\delta$ and $\delta_{\text{m}}$ represent isotropic deviation of the Hubble parameter and the matter overdensity respectively. Here, $H(t)$ and $\rho(t)$ represent the zero order quantities, and hence satisfy Eqs. \eqref{eq:00-zero}, \eqref{eq:trace-zero} and \eqref{eq:continuity-zero} (in some references, these are sometimes denoted as $H_0(t)$ and $\rho_0(t)$, however such notation is avoided here to easily distinguish from quantities which are evaluated at present times).

The perturbation of the function $f$ and its derivatives are
\begin{equation}
\delta f = f_T \delta T, \: \delta f_T = f_{TT} \delta T,
\end{equation}
where $\delta x$ represents the first-order perturbation of the variable $x$. Here, $\delta T = 2T \delta$. In this way, the perturbed equations of Eqs. \eqref{eq:00-zero} and \eqref{eq:continuity-zero} become
\begin{align}
-T\left(1+f_T-12H^2f_{TT}\right)\delta &= \kappa^2 \rho \delta_{\text{m}}, \label{eq:00-first} \\
\dot{\delta}_{\text{m}} + 3H(1+w)\delta &= 0. \label{eq:continuity-first}
\end{align}

The relationship between $\delta$ and $\delta_{\text{m}}$ can be expressed in terms of $T$ by using Eq. \eqref{eq:00-zero} in Eq. \eqref{eq:00-first} to give
\begin{equation}\label{eq:delta-deltam-rel}
\delta = \dfrac{1}{2T}\dfrac{T+2Tf_T-f}{1+f_T+2Tf_{TT}}\delta_{\text{m}}.
\end{equation}
In this way, an expression for $\delta_{\text{m}}$ can be found by substituting the previous relationship in Eq. \eqref{eq:continuity-first} 
\begin{equation}
\dot{\delta}_{\text{m}} + \dfrac{3H}{2T}(1+w)\dfrac{T+2Tf_T-f}{1+f_T+2Tf_{TT}}\delta_{\text{m}}= 0,
\end{equation}
which is a separable first-order ODE in $\delta_{\text{m}}$, whose solution is given by
\begin{equation}\label{eq:deltam-sol1}
\delta_{\text{m}} = \exp\left[-\dfrac{3}{2}(1+w)\int \dfrac{H}{T}\dfrac{T+2Tf_T-f}{1+f_T+2Tf_{TT}} dt\right].
\end{equation}

The integral can be solved analytically as follows. Using the continuity equation \eqref{eq:continuity-zero} for a perfect fluid and using the zeroth order $tt$-component of the Friedmann equations Eq. \eqref{eq:00-zero} with Eq. \eqref{eq:trace-zero} gives
\begin{equation}
\dot{H} = \dfrac{1}{4}(1+w) \dfrac{T+2T f_T -f}{1 + f_T + 2T f_{TT}}.
\end{equation}
Substituting in Eq. \eqref{eq:deltam-sol1} yields
\begin{equation}\label{eq:deltam-solution}
\delta_{\text{m}} = \exp\left[\int \dfrac{\dot{H}}{H} dt\right] = \exp\left[\int \dfrac{dH}{H}\right] = kH,
\end{equation}
where $k$ is a constant of integration, which can be determined by evaluating $\delta_{\text{m}}$ at current times. This in turn gives $k = \delta_{\text{m}}\left(t_0\right)/H_0$. Therefore, using Eq. \eqref{eq:delta-deltam-rel}, $\delta$ is found to be
\begin{equation}
\delta = -\dfrac{\delta_{\text{m}}\left(t_0\right)}{3(1+w)H_0} \dfrac{\dot{H}}{H}.
\end{equation}
Depending on the current value of $\delta_{\text{m}}$, the evolutions of both $\delta$ and $\delta_{\text{m}}$ change. One further notes that $w = -1$ poses a singularity for $\delta$. Furthermore, the solution of $\delta$ can be expressed in terms of its current value as
\begin{equation}\label{eq:delta-solution}
\delta = \delta\left(t_0\right) \dfrac{u(t)}{u(t_0)},
\end{equation}
where $u(t) \equiv \dfrac{\dot{H}}{H}$, $\delta\left(t_0\right) \equiv -\dfrac{\delta_{\text{m}}\left(t_0\right)}{3(1+w)H_0}  \dfrac{\dot{H}}{H}\bigg|_{t = t_0}$.

Stability is achieved as long as both $\delta_{\text{m}}$ and $\delta$ decay with cosmic time. As an order approximation, this seems to be the case for both $\delta$ and $\delta_{\text{m}}$ since $\delta \sim \delta_{\text{m}} \sim t^{-1}$ i.e. decays with time. Therefore we investigate various $f(T)$ functions to determine their stability and whether such claim holds.

\section{IV. Power-law and de-Sitter stability}

\subsection{A. Power-law stability}

Consider a power-law type evolution for the scale factor, i.e. $a(t) \propto t^m$ where $m$ is a constant. The Hubble parameter in this case becomes $H(t) = \dfrac{m}{t}$ and therefore $T = -6\dfrac{m^2}{t^2}$. Hence, the perturbation variables $\delta$ and $\delta_{\text{m}}$ take the form
\begin{equation}
\delta = \delta\left(t_0\right) \dfrac{t_0}{t}, \: \delta_{\text{m}} = \delta_{\text{m}}\left(t_0\right) \dfrac{t_0}{t}.
\end{equation}
Given the inverse relationship with $t$, both perturbations die out for late times. However, since a specific form of $a(t)$ is considered, only specific functions of $f(T)$ are allowed. Such classes of functions can be found using the $tt$-component of the Friedmann equations Eq. \eqref{eq:00-zero}. From the definition of $T$, one finds
\begin{equation}\label{eq:tT_power_rel}
\dfrac{t}{t_0} = \left(\dfrac{T}{T_0}\right)^{-1/2}.
\end{equation}
On the other hand, $a(t)$ can be expressed as
\begin{equation}
a(t) = \left(\dfrac{t}{t_0}\right)^m.
\end{equation}
Substituting these two equations and Eq. \eqref{eq:continuity-zero-solution} into Eq. \eqref{eq:00-zero} yields
\begin{equation}
1+2f_T -\dfrac{f}{T} = \Omega_{w,0} \left(\dfrac{T}{T_0}\right)^{\frac{3m(1+w)}{2}-1},
\end{equation}
where $\Omega_{w,0}$ is the current value of the density parameter of the fluid having EoS parameter $w$. This is a first order differential equation in $f(T)$ which has two solutions depending on the nature of the power.

\subsubsection{I. $-1+3 m (w+1) \neq 0$}

For the case when $-1+3 m (w+1) \neq 0$, the solution is given by
\begin{equation}
f(T) = -T+c_1 \sqrt{-T}+\dfrac{\Omega_{w,0} T_0 \left(\frac{T}{T_0}\right)^{\frac{3}{2} m (w+1)}}{-1+3 m (w+1)}
\end{equation}
where $c_1$ is an integration constant. The exotic fluid's energy density is then given by
\begin{equation}\label{eq:rhoDE_power-law_case1}
2\kappa^2 \rho_{\text{exo}} = -T + \Omega_{w,0}\left(\frac{T}{T_0}\right)^{3m (1 + w)/2} T_0.
\end{equation}
One notes that the effect of the square root term does not contribute neither to the Friedmann equations nor to the energy density. For this reason, this term is neglected by setting the integration constant to zero. The Lagrangian thus reduces to
\begin{equation}
\mathcal{L}_{\text{grav}} = \dfrac{\Omega_{w,0} T_0 \left(\frac{T}{T_0}\right)^{\frac{3}{2} m (w+1)}}{-1+3 m (w+1)}.
\end{equation}
In the case of GR, standard power-law solutions are obtained when the power of $T$ equals $1$, i.e. $\frac{3}{2} m (w+1) = 1$. For example, in matter only universes, $w = 0$ and $m = 2/3$ whilst in radiation only universes, $w = 1/3$ and $m = 1/2$. However, we are interested in the non-trivial solutions. For the solution to have physical meaning, we first require the energy density Eq. \eqref{eq:rhoDE_power-law_case1} to be positive.  This leads to the following constraint,
\begin{equation}
\Omega_{w,0} \leq \left(\frac{T}{T_0}\right)^{1-\frac{3m (1 + w)}{2}} = \left(\frac{t}{t_0}\right)^{-2+3m (1 + w)}
\end{equation}
where Eq. \eqref{eq:tT_power_rel} was used. Since $\Omega_{w,0}$ is a constant in cosmic time, the inequality only holds for two different scenarios: 

\paragraph{Case 1} if $-2+3m (1 + w) = 0$, then $\Omega_{w,0} \leq 1$. This is the rescaling of GR case (the Lagrangian is $\mathcal{L}_{\text{grav}} = \Omega_{w,0} T$).
\paragraph{Case 2} if $-2+3m (1 + w) \neq 0$, $\Omega_{w,0} = 0$, which is non-physical.

Therefore, non-trivial power-law solutions (except for GR rescaling) are not possible when $-1+3 m (w+1) \neq 0$. 

\subsubsection{II. $-1+3 m (w+1) = 0$}

In this case, the solution for $f(T)$ is given by,
\begin{equation}
f(T) = -T + c_1 \sqrt{-T}+\frac{1}{2} \Omega_{w,0} T_0 \sqrt{\frac{T}{T_0}} \ln \left(\frac{T}{T_0}\right),
\end{equation}
where $c_1$ is an integration constant. The exotic fluid's energy density is 
\begin{equation}
2\kappa^2 \rho_{\text{exo}} = -T + \Omega_{w,0} \sqrt{\frac{T}{T_0}} T_0.
\end{equation}
Once again the effect of the square root term in both the field equation and energy density does not contribute, and therefore is neglected. This reduces the Lagrangian to
\begin{equation}
\mathcal{L}_{\text{grav}} = \frac{1}{2} \Omega_{w,0} T_0 \sqrt{\frac{T}{T_0}} \ln \left(\frac{T}{T_0}\right).
\end{equation}
In this scenario, the action cannot reduce to GR so it is in a sense non-trivial. However, the solution can only be physical provided that the energy density is positive. This leads to the condition, 
\begin{equation}
\Omega_{w,0} \leq \left(\frac{T}{T_0}\right)^{1/2} = \frac{t_0}{t},
\end{equation}
where again Eq. \eqref{eq:tT_power_rel} was used. Since $\Omega_{w,0}$ is a constant in cosmic time, the inequality only holds if $\Omega_{w,0} = 0$, which is non-physical. 

Therefore, power-law solutions can only be achieved under GR (or rescaling) and such solutions are stable, as expected. 

\subsection{B. de-Sitter stability}

A de-Sitter universe is achieved when the Hubble parameter becomes constant, i.e. $H(t) = H_0$, which leads to $T = -6 {H_0}^2 = T_0$. This induces an expansion history for $a(t)$ of the form
\begin{equation}
a(t) = e^{H_0 \left(t-t_0\right)}.
\end{equation}
The perturbation variables $\delta$ and $\delta_{\text{m}}$ reduce to
\begin{equation}
\delta = 0, \: \delta_{\text{m}}(t) = \delta_{\text{m}}\left(t_0\right).
\end{equation}
This shows that the solution is stable, being perfectly homogeneous but having a constant isotropic perturbation (which occurs also in $\Lambda$CDM models during late times). However, as in the previous section, since a specific form of $a(t)$ is considered, only specific functions of $f(T)$ are allowed. The set of permissible functions can be found using the $tt$-component of the Friedmann equations Eq. \eqref{eq:00-zero}, which yields
\begin{equation}
\left(1+2f_T-\dfrac{f}{T}\right)\bigg|_{T = T_0} = \Omega_{w,0} e^{-3(1+w)H_0 \left(t-t_0\right)},
\end{equation}
where  Eq. \eqref{eq:continuity-zero-solution} was used. However, the Friedmann equation poses a possible contradiction, the left hand side (LHS) is a constant in time whilst the right hand side (RHS) is time dependent. The only possible scenario at which this situation is avoided is if $w = -1$ (in other words, a cosmological constant type fluid must exist which is similar to GR where de-Sitter evolution is obtained when only the cosmological constant is present). In this special case, the Friedmann equation becomes
\begin{equation}
\left(1+2f_T-\dfrac{f}{T}\right)\bigg|_{T = T_0} = \Omega_{w,0}. 
\end{equation}
This can be treated as a first order ODE in $T_0$, whose solution is given by
\begin{equation}
f\left(T=T_0\right) = \left(\Omega_{w,0}-1\right) T_0 + c_1 \sqrt{-T_0}
\end{equation}
where $c_1$ is an integration constant. Similar to the power-law case, the square root term does not play a role neither in the Friedmann equations nor in the exotic fluid energy density, and hence is not included in the Lagrangian being considered. Therefore, the Lagrangian becomes
\begin{equation}
\mathcal{L}_{\text{grav}} = \Omega_{w,0} T_0.
\end{equation}
Such a Lagrangian is a rescaling of the standard TEGR Lagrangian, as obtained in the power-law case. This should not deviate much from TEGR, i.e $\Omega_{w,0} \approx 1$. Nonetheless, this is essentially GR (the only difference lies in the definition of the Newtonian gravitational constant) and not a true new non-trivial solution for de-Sitter cosmology. 

Note that in this case,the exotic fluid energy density is given by
\begin{equation}\label{eq:rhoDE_de-Sitter}
2\kappa^2 \rho_{\text{exo}} = -T_0\left[1- \Omega_{w,0} e^{-3(1+w)H_0 \left(t-t_0\right)}\right],
\end{equation}
which for $w = -1$ simply reduces to $2\kappa^2 \rho_{\text{exo}} = -T_0\left[1- \Omega_{w,0}\right]$. This is physical only when the energy density is positive, in other words when $1- \Omega_{w,0} \geq 0 \implies \Omega_{w,0} \leq 1$.

\section{V. Stability for different $f(T)$ ansatz}

\subsection{A. $f(T) = \alpha (-T)^n$ stability}

In this section, we consider the power-law ansatz model, originally proposed by Bengochea and Ferraro \cite{Bengochea:2008gz}, which is of the form $f(T) = \alpha (-T)^n$, where $\alpha$ and $n$ are constants. Furthermore, for simplicity, we assume that the matter density is solely composed of matter (hence $w = 0$). Thus, Eq. \eqref{eq:00-zero} becomes
\begin{equation}\label{eq:tt-power-law}
\alpha (2n-1) \dfrac{(-T)^n}{T_0}+\dfrac{T}{T_0}= \dfrac{\Omega_{\text{M},0}}{a(t)^3},
\end{equation}
The value of $\alpha$ can be found by evaluating the expression at current times, to give
\begin{equation}
\alpha = \dfrac{(\Omega_{\text{M},0}-1) \left(6 {H_0}^2\right)^{1-n}}{1-2 n},
\end{equation}
where $\Omega_{\text{M},0}$ is the matter density today. Note that this do not hold for $n = 1/2$, as such power does not contribute to the Friedmann equation (see Eq. \eqref{eq:tt-power-law}). Using this value of $\alpha$, the equation reduces further into
\begin{equation}
\dfrac{T}{T_0}+(\Omega_{\text{M},0}-1) \left(\dfrac{T}{T_0}\right)^{n}=\dfrac{\Omega_{\text{M},0}}{a(t)^3}.
\end{equation}
In this case, the energy density for the exotic fluid is expressed as,
\begin{equation}
2\kappa^2 \rho_{\text{exo}} = (\Omega_{\text{M},0}-1) T_0 \left(\dfrac{T}{T_0}\right)^n.
\end{equation}
Note that this directly implies that $\rho_{\text{exo}} \geq 0$ for every $n$ and hence this function can lead to physical solutions. The exotic fluid's density is thus found to be
\begin{equation}
\Omega_T (t) \equiv \dfrac{\rho_{\text{exo}}}{\rho_{cr}} = (1-\Omega_{\text{M},0}) \left(\dfrac{T}{T_0}\right)^n.
\end{equation}

Using Eq. \eqref{eq:def-EoS-DE}, the equation of state parameter, $w_{\text{exo}}$, becomes
\begin{equation}
w_{\text{exo}} = \dfrac{(n-1)}{1+n (\Omega_{\text{M},0}-1) \left(\frac{T}{T_0}\right)^{n-1}}.
\end{equation}
For $n = 0$, this reduces to the standard $\Lambda$CDM model (with $w_{\text{exo}} = -1$ as expected). For $n \leq 0$, the EoS parameter is negative at all times. This is also achieved for $0 < n < 1$ (except for $n = 1/2$), which can be seen by using the $tt$-Friedmann equation Eq. \eqref{eq:00-zero} to express $w_{\text{exo}}$ as
\begin{equation}
w_{\text{exo}} = -\left[1+\dfrac{n}{1-n}\dfrac{\Omega_{\text{M},0}}{a(t)^3}\dfrac{T_0}{T}\right]^{-1}.
\end{equation}
At current times, the EoS parameter is given by
\begin{equation}
w_{\text{exo}}(t_0) = -\left[1+\dfrac{n}{1-n}\Omega_{\text{M},0}\right]^{-1}.
\end{equation}
According to recent PLANCK data, constant EoS dark energy parameter models (i.e. $\Lambda$CDM and $w$CDM) and linear evolution EoS dark energy parameter models [e.g. Chevallier-Polarski-Linder (CPL) \cite{Chevallier:2000qy,Linder:2002et}] indicate a current value of approximately $-1$ \cite{Kumar:2012gr,Capozziello:2013wha,Aubourg:2014yra,Ade:2013zuv,Magana:2014voa}. However, such models are not directly comparable with the model considered here since the evolution is fundamentally different. Nonetheless, since the models seem to indicate that at current times that the exotic fluid's EoS parameter is approximately $-1$, we shall focus on models which exhibit a similar feature. 

The true constraint on the power $n$ however is attributed to the fact that the universe is observed to be accelerating today. This is achieved by the decelerating parameter, which for power-law models is given by
\begin{equation}
q(t) = \frac{1-(2 n-3) (\Omega_{\text{M},0}-1) \left(\dfrac{T}{T_0}\right)^{n-1}}{2+2 n (\Omega_{\text{M},0}-1)  \left(\dfrac{T}{T_0}\right)^{n-1}}.
\end{equation}
Since acceleration is observed at current times, one requires that $q\left(t_0\right) < 0$. In this model, the deceleration parameter at current times is given by
\begin{equation}
q\left(t_0\right) = \frac{1-(2 n-3) (\Omega_{\text{M},0}-1)}{2+2 n (\Omega_{\text{M},0}-1)}.
\end{equation} 
By assuming that $\Omega_{\text{M},0} \approx 0.3$ leads to $n<0.75$ or $n>2.18$. Since we also require the exotic fluid's EoS parameter to be close to $-1$, this allows us to consider two power-law possibilities, being $n = -1$ ($w_{\text{exo}} \approx -1.18$) and $n = -2$ ($w_{\text{exo}} = -1.25$). 

In the following figures, we analyse some of the features for power-law cosmologies with $n = 0$ (i.e. $\Lambda$CDM), $n = -1$ and $n = -2$. The plots were carried out using $\Omega_{\text{M},0} = 0.3$ and $H_0 = \left(\SI{14.4}{\giga\year}\right)^{-1}$. The solid curve represents $\Lambda$CDM, whilst the dotted and dashed curves represent $n = -1$ and $n = -2$ respectively. 

\begin{figure}[h!]
\includegraphics[width=0.49\textwidth{}]{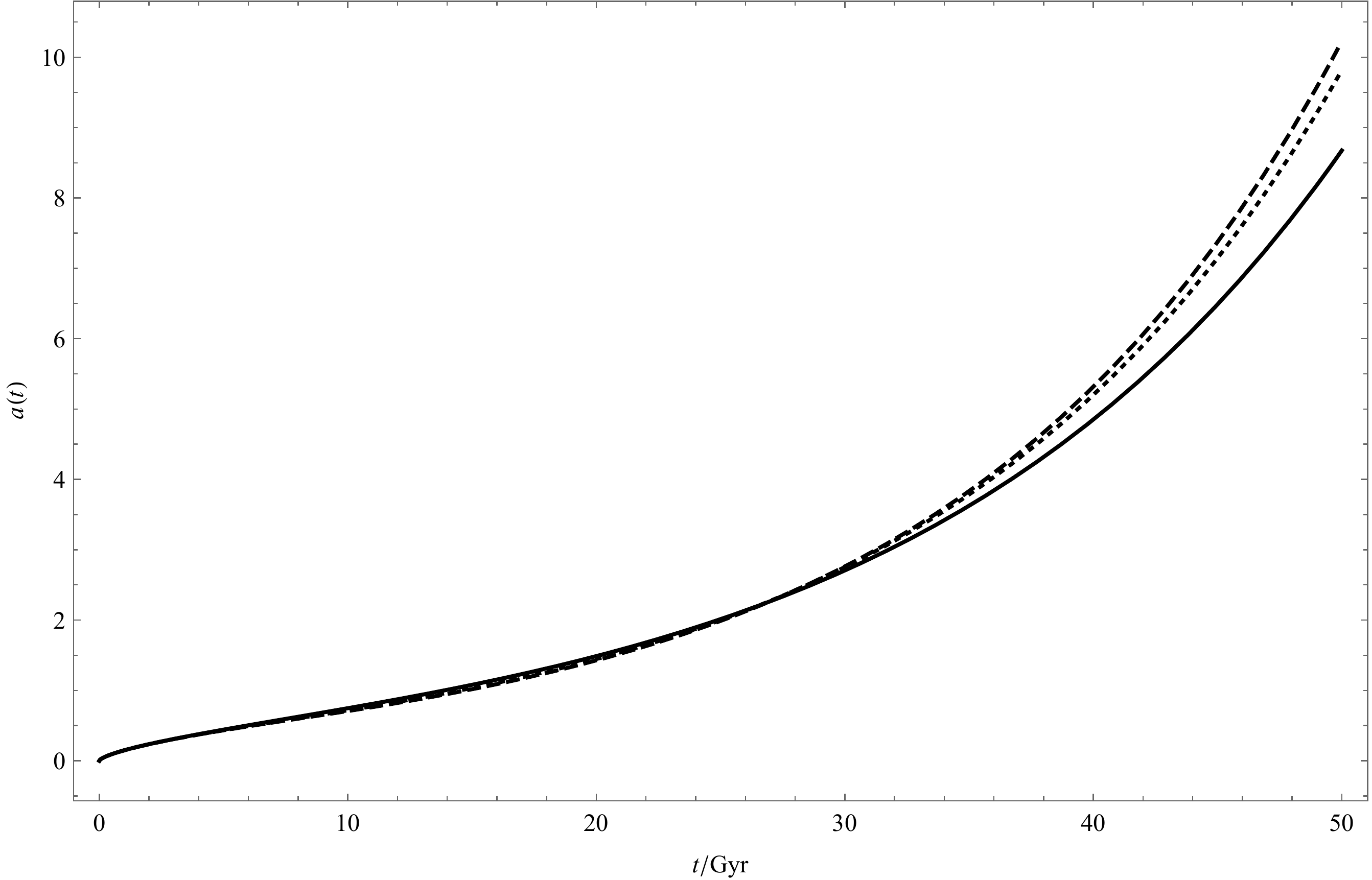}
\caption{The evolution of the scale factor $a(t)$ with cosmic time for the model $f(T) = \alpha \left(-T\right)^n$ for $n = 0, -1$ and $-2$. For early and current times, the scale factor for each model is essentially identical. On the other hand, during late times, the $n = -1$ (dotted) and $n = -2$ (dashed) models start to deviate from the $\Lambda$CDM (solid) model.}
\label{fig:fig_scalefactor_power}
\end{figure}

We start by first analysing the evolution of the scale factor with cosmic time, shown in Fig. \ref{fig:fig_scalefactor_power}. One notes that the $n = -1$ and $n = -2$ only differ from $\Lambda$CDM at late times, while the two do not differ by much from each other.

\begin{figure}[h!]
\includegraphics[width=0.49\textwidth{}]{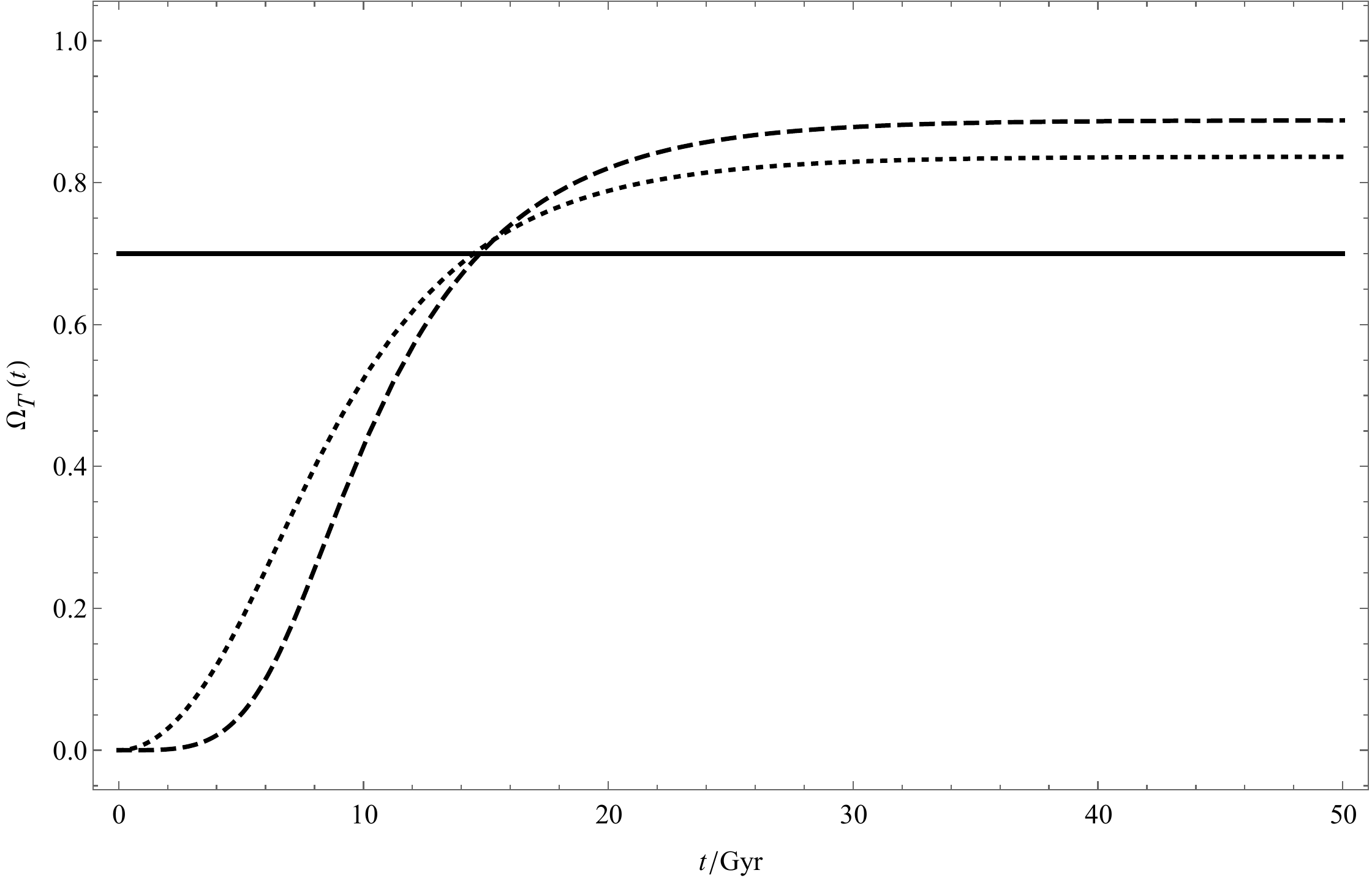}
\caption{The evolution of the torsional density parameter $\Omega_T(t)$ with cosmic time for the model $f(T) = \alpha \left(-T\right)^n$ for $n = 0, -1$ and $-2$. In contrast to $\Lambda$CDM (solid) which sits at a constant value of 0.7, the $n = -1$ (dotted) and $n = -2$ (dashed) models initially start from a value of 0, increases until it reaches 0.7 at current times and keeps on increases until it reaches a limiting constant value at late times [$\Omega_T \left(t = t_\infty\right) \approx 0.85$ for $n = -1$ and $\Omega_T \left(t = t_\infty\right) \approx 0.90$ for $n = -2$].}
\label{fig:fig_OmegaT_power}
\end{figure}

Fig. \ref{fig:fig_OmegaT_power} shows the variation of $\Omega_T$ with cosmic time. The models vary greatly initially from $\Lambda$CDM simply because the $\Lambda$CDM model has a constant dark energy density and hence stays constant in time, whilst in this case the exotic fluid's energy density changes with time. However, it retains a common feature with $\Lambda$CDM in which at late times, the density parameter becomes constant [$\Omega_T \left(t = t_\infty\right) \approx 0.85$ for $n = -1$ and $\Omega_T \left(t = t_\infty\right) \approx 0.90$ for $n = -2$].

\begin{figure}[h!]
\includegraphics[width=0.49\textwidth{}]{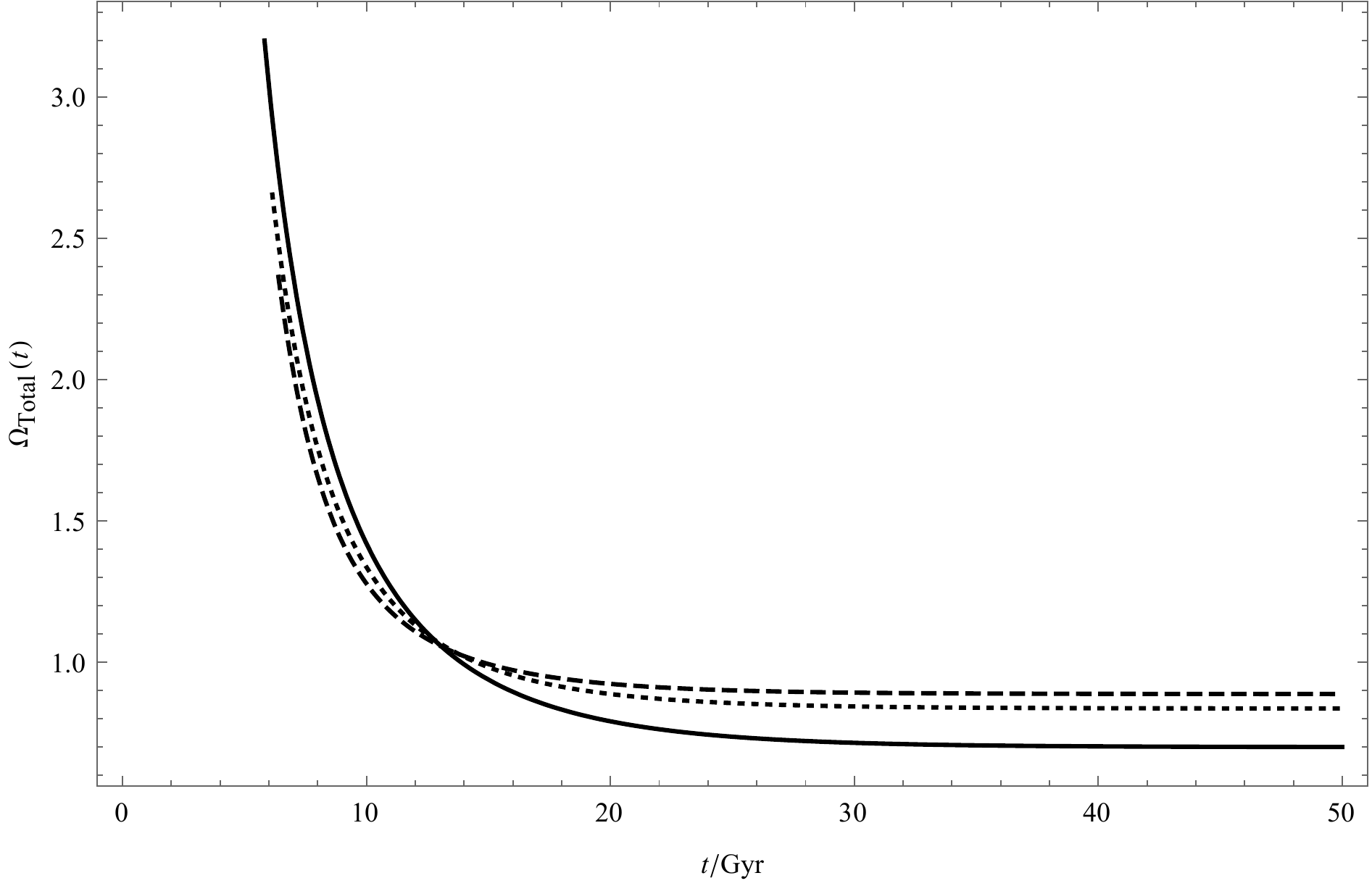}
\caption{The evolution of the total density parameter $\Omega_{\text{Total}}(t)$ with cosmic time for the model $f(T) = \alpha \left(-T\right)^n$ for $n = 0, -1$ and $-2$. One notes that all models exhibit a similar evolution behaviour in which it starts at very high values due to the matter density being dominant and the torsional component being irrelevant, reaches 1 at today's time and decreases to a constant value at late times given by the torsional density (since the matter density becomes irrelevant). The only difference between the theories lies in the late time values (for $\Lambda$CDM (solid) it is equal to 0.7 whilst for $n = -1$ (dotted) and $n = -2$ (dashed) are 0.85 and 0.90 respectively).}
\label{fig:fig_OmegaTotal_power}
\end{figure}

Through the variation of $\Omega_T$, one can also analyse the behaviour of the total energy density parameter $\Omega_{\text{Total}} \equiv \Omega_\text{M} + \Omega_T$. This is described in Fig. \ref{fig:fig_OmegaTotal_power}, in which one finds that the power-law models considered exhibit the same behaviour as $\Lambda$CDM, with the only major difference occurring during late times. Each model has a limiting value during late times, which occurs since the effect becomes irrelevant $\left[\Omega_{\text{M}} \propto a(t)^{-3}\right]$ and $\Omega_T$ limits to a constant value (in the case of $\Lambda$CDM, this is equal to 0.7).  For the power-law models considered, the limiting values are 0.85 and 0.90 for $n = -1$ and $n = -2$ respectively.

\begin{figure}[h!]
\includegraphics[width=0.49\textwidth{}]{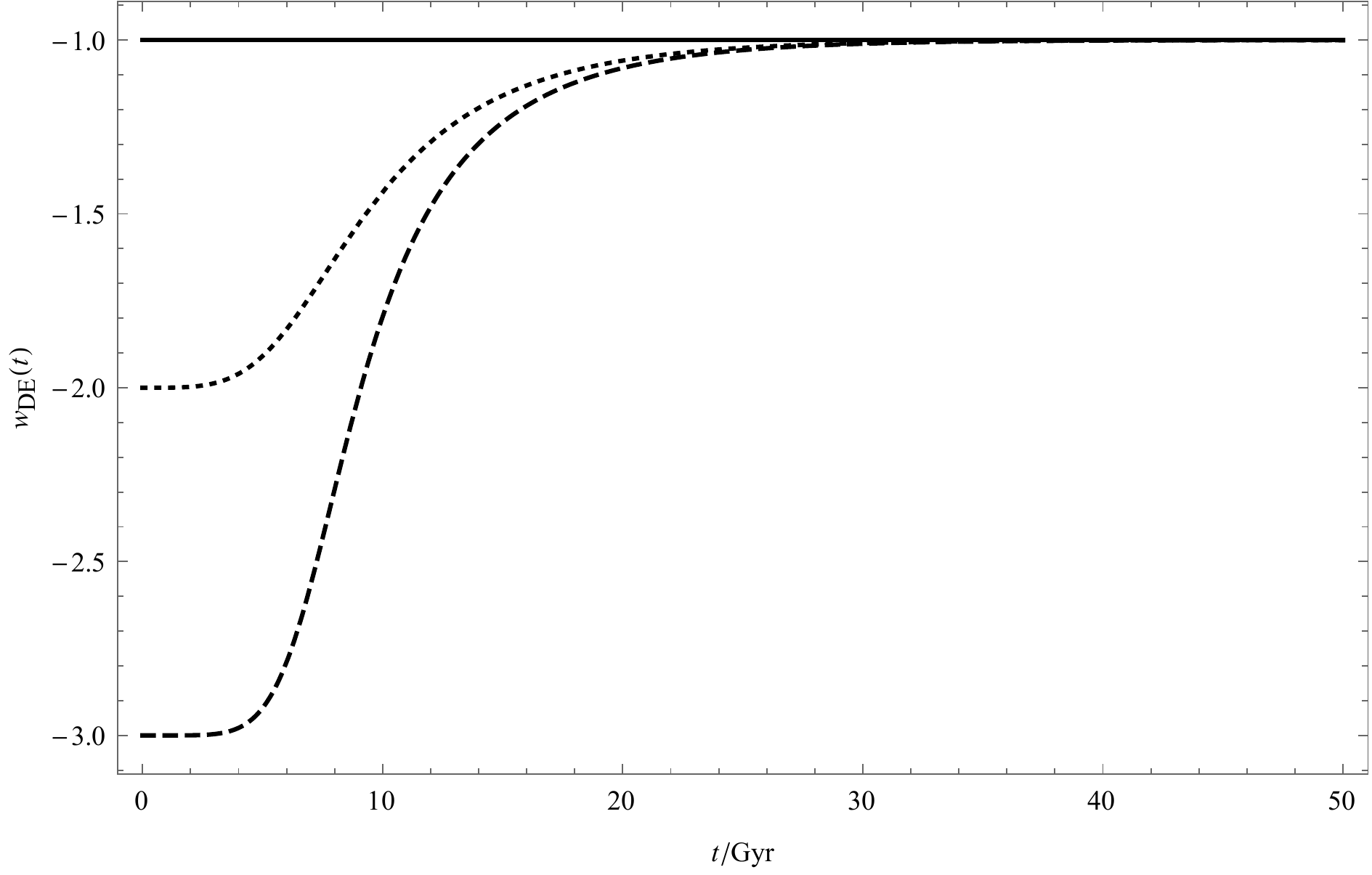}
\caption{The evolution of the exotic fluid's EoS parameter $w_{\text{exo}}(t)$ with cosmic time for the model $f(T) = \alpha \left(-T\right)^n$ for $n = 0, -1$ and $-2$. In $\Lambda$CDM (solid), the EoS parameter is constant and equals to $-1$ (since this describes a cosmological constant). On the other hand, the $n = -1$ (dotted) and $n = -2$ (dashed) models describe a varying EoS parameter, starting at $n - 1$ at $t = 0$ and reaches $-1$ during late times. In other words, such models mimic the cosmological constant behaviour at late times. For all cases, the nature of the exotic fluid is phantom.}
\label{fig:fig_wDE_power}
\end{figure}

The variation of the EoS parameter $w_{\text{exo}}$ is shown in Fig. \ref{fig:fig_wDE_power}. Again, since the exotic fluid's energy density is not constant this leads to a variation of the EoS parameter. One notes that the parameter varies from $n-1$ initially and approaches $-1$ at late times, i.e. effectively becoming a cosmological constant in nature at late times. Furthermore, the plot infers that for such models, the exotic fluid is phantom in nature. 

\begin{figure}[h!]
\includegraphics[width=0.49\textwidth{}]{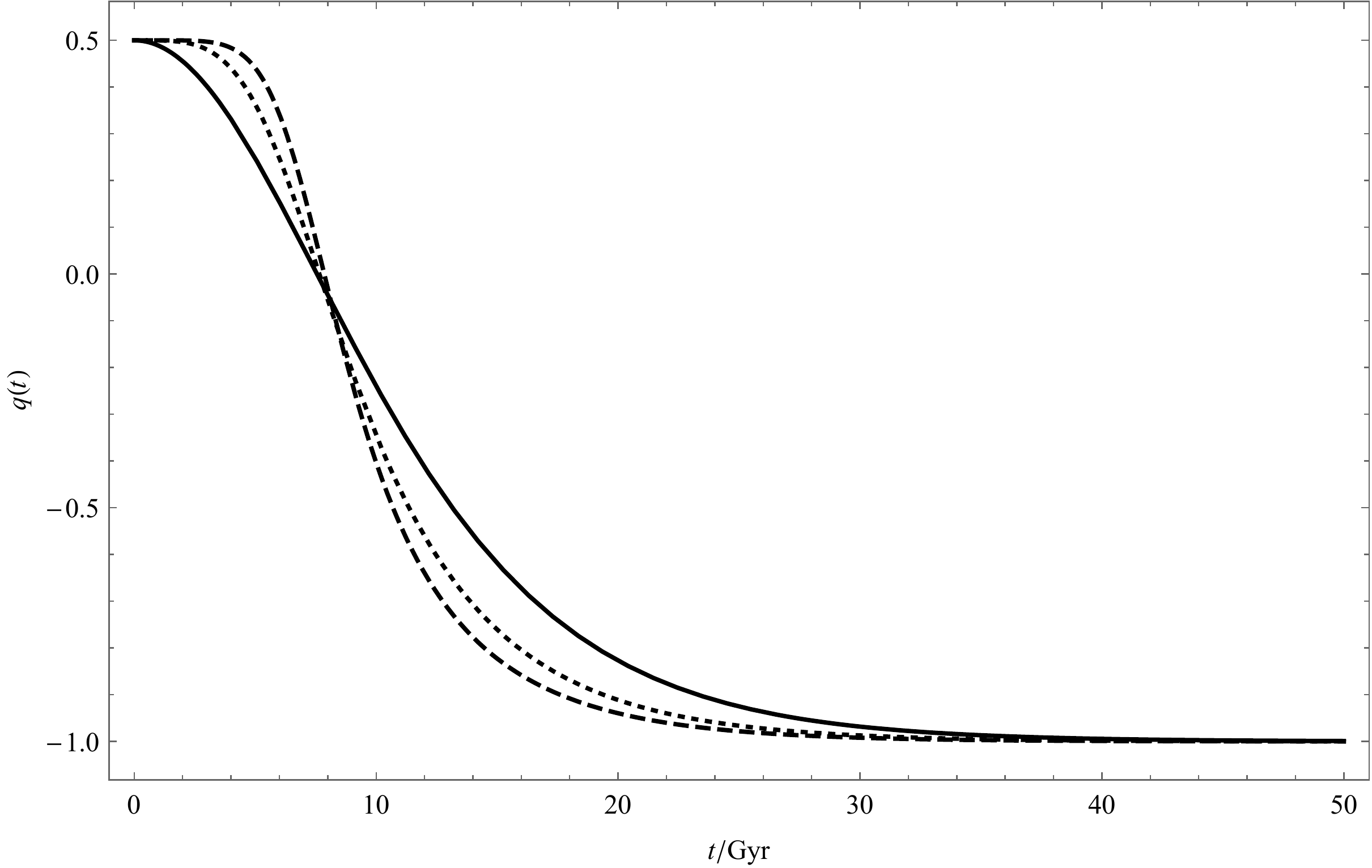}
\caption{The evolution of the deceleration parameter $q(t)$ with cosmic time for the model $f(T) = \alpha \left(-T\right)^n$ for $n = 0, -1$ and $-2$. All models mimic the behaviour of the $\Lambda$CDM (solid) model, and each of them transition to an accelerating universe at approximately the same time $(t \approx \SI{8}{\giga\year})$. Each model starts with $q(0) = 0.4$ and ends with $q\left(t = t_\infty\right) = -1$ as values. What can be noticed however is that the rate of $q(t)$ happens much more drastically for $n = -1$ (dotted) and $n = -2$ (dashed) compared to $\Lambda$CDM.}
\label{fig:fig_dec_power}
\end{figure}

In Fig. \ref{fig:fig_dec_power}, one sees the variation of the deceleration parameter with time. One notes that each model follows a $\Lambda$CDM-like plot, with the transition to an accelerated expansion starting around the same period. Since each model reaches $w_{\text{exo}} \rightarrow -1$ at late times, the deceleration parameter $q$ approaches $-1$ as well. The transition between non-accelerating and accelerating occurs at approximately the same time $(t \approx \SI{8}{\giga\year})$.

\begin{figure}[h!]
\includegraphics[width=0.49\textwidth{}]{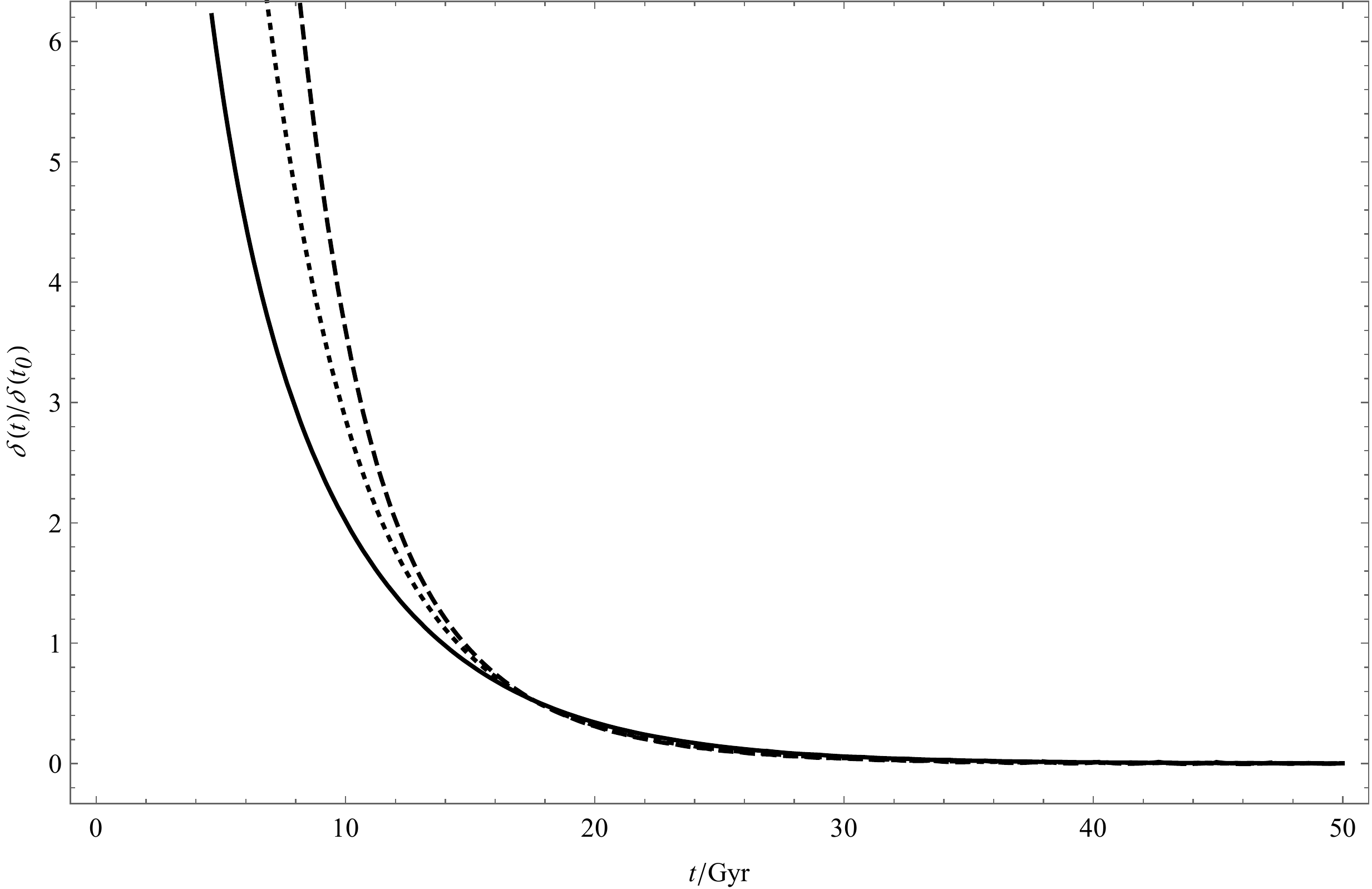}
\caption{The evolution of the ratio of the Hubble perturbation parameter $\delta(t)$ to its current value $\delta\left(t_0\right)$ with cosmic time for the model $f(T) = \alpha \left(-T\right)^n$ for $n = 0, -1$ and $-2$. Each model exhibits the same behaviour as $\Lambda$CDM, in which each model decays to 0 at late times indicating stability. The difference only occurs during earlier times where the value of $\delta(t)$ varies from model to model.}
\label{fig:fig_delta_power}
\end{figure}

\begin{figure}[h!]
\includegraphics[width=0.49\textwidth{}]{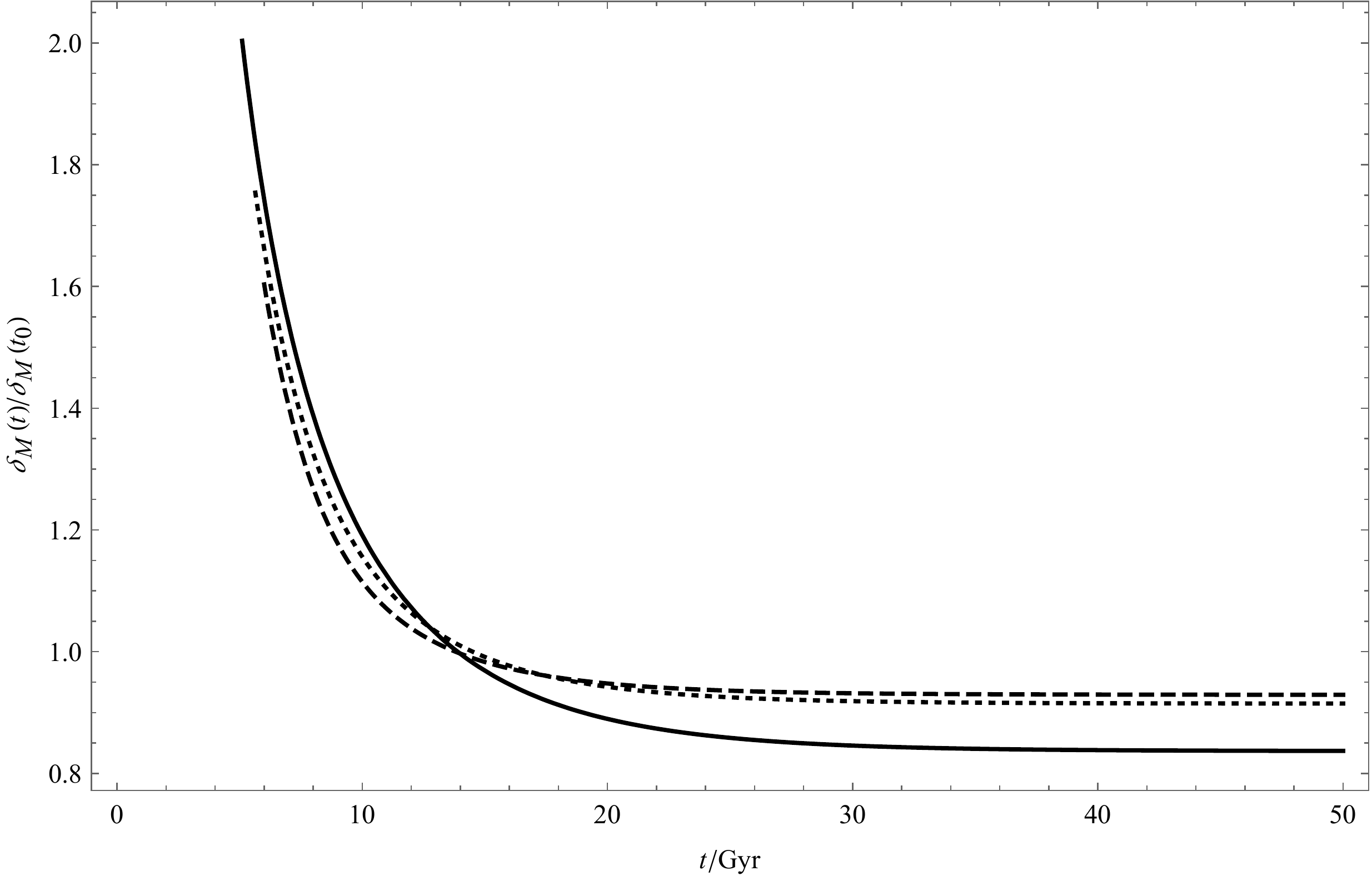}
\caption{The evolution of the ratio of the matter perturbation parameter $\delta_\text{M}(t)$ to its current value $\delta_\text{M} \left(t_0\right)$ with cosmic time for the model $f(T) = \alpha \left(-T\right)^n$ for $n = 0, -1$ and $-2$. One again finds that each model evolves in the same way as $\Lambda$CDM, in which each of them approach the same value of $1$ at current times and decays to a constant value at late times. This value is the only major difference between the models [for $\Lambda$CDM (solid), $\delta(t)/\delta\left(t_0\right) \approx 0.850$ whilst for $n = -1$ (dotted) and $n = -2$ (dashed), $\delta(t)/\delta\left(t_0\right) \approx 0.925$]. Furthermore, each model is stable and indicate individually that the matter perturbations persist at late times.}
\label{fig:fig_deltaM_power}
\end{figure}

Figs. \ref{fig:fig_delta_power} and \ref{fig:fig_deltaM_power} describe the evolution of the perturbations. For the perturbations of the Hubble parameter (Fig. \ref{fig:fig_delta_power}), the solutions indicate stability, and decay in the same way as $\Lambda$CDM models, and completely decay at late times. In the case for matter perturbations, the models again exhibit a similar behaviour as $\Lambda$CDM in that they are stable and decaying with time until they reach a limiting value at late times. This means that the matter perturbations persist at late times in the same way as $\Lambda$CDM does with the only difference being the limiting values [for $\Lambda$CDM, $\delta(t)/\delta\left(t_0\right) \approx 0.850$ whilst for $n = -1$ and $n = -2$, $\delta(t)/\delta\left(t_0\right) \approx 0.925$].

\subsection{B. $f(T) = \alpha T_0 \left(1 - \exp \left[-p \sqrt{\dfrac{T}{T_0}}\right]\right)$ stability}

In this section, we consider Linder's exponential gravity model for $f(T)$ which is of the form $f(T) = \alpha T_0 \left(1 - \exp \left[-p \sqrt{\dfrac{T}{T_0}}\right]\right)$, where $\alpha$ and $p$ are constants with $p \neq 0$ (since $p = 0$ leads to $f(T) = 0$, i.e. TEGR) \cite{Linder:2010py}. Similar to the previous section, for simplicity, we again assume that the matter density is solely composed of matter (giving $w = 0$). Thus, Eq. \eqref{eq:00-zero} becomes
\begin{align}
&\dfrac{T}{T_0}-\alpha \left\lbrace 1-\left(1+p\sqrt{\dfrac{T}{T_0}}\right)\exp \left[-p \sqrt{\dfrac{T}{T_0}}\right]\right\rbrace \nonumber \\
&\hspace{1cm} = \dfrac{\Omega_{\text{M},0}}{a(t)^3},
\end{align}
The value of $\alpha$ can be found by evaluating the expression at current times, to give
\begin{equation}
\alpha = -\frac{\left(\Omega_{\text{M},0}-1\right)}{1-\left(1+p\right)e^{-p}},
\end{equation}
where $\Omega_{\text{M},0}$ is the matter density today. Using this value of $\alpha$, the equation reduces further into
\begin{align}
\dfrac{\Omega_{\text{M},0}}{a(t)^3} &= \dfrac{T}{T_0}+\frac{\left(\Omega_{\text{M},0}-1\right)}{1-\left(1+p\right)e^{-p}} \bigg\lbrace 1-\left(1+p\sqrt{\dfrac{T}{T_0}}\right) \nonumber \\
&\times \exp \left[-p \sqrt{\dfrac{T}{T_0}}\right]\bigg\rbrace,
\end{align}
For this model, the exotic fluid's energy density is given by
\begin{align}
2\kappa^2 \rho_{\text{exo}} &= \dfrac{\left(\Omega_{\text{M},0}-1\right) T_0}{1-\left(1+p\right)e^{-p}}\bigg\lbrace 1-\left(1+p\sqrt{\dfrac{T}{T_0}}\right) \nonumber \\
&\times \exp \left[-p \sqrt{\dfrac{T}{T_0}}\right]\bigg\rbrace.
\end{align}
One notes that the curly bracketed term and the denominator are always positive for $p \neq 0$. Since the product $\left(\Omega_{\text{M},0}-1\right) T_0$ is also positive, then the energy density becomes always positive, and hence can lead to physical solutions. Thus, the exotic fluid's density parameter is found to be
\resizebox{0.99\linewidth}{!}{
  \begin{minipage}{\linewidth}
\begin{align}
&\Omega_T (t) \equiv \dfrac{\rho_{\text{exo}}}{\rho_{\text{cr}}} \nonumber \\
&= \dfrac{\left(1-\Omega_{\text{M},0}\right) \left\lbrace 1-\left(1+p\sqrt{\dfrac{T}{T_0}}\right)\exp \left[-p \sqrt{\dfrac{T}{T_0}}\right]\right\rbrace}{1-\left(1+p\right)e^{-p}}.
\end{align}
  \end{minipage}
}

The EoS parameter $w_{\text{exo}}$ for this model becomes
\begin{equation}
w_{\text{exo}} = -\frac{2-\dfrac{p^2 \dfrac{T}{T_0}\exp \left[-p \sqrt{\dfrac{T}{T_0}}\right]}{1-\left(1+p  \sqrt{\dfrac{T}{T_0}}\right)\exp \left[-p \sqrt{\dfrac{T}{T_0}}\right]}}{2+\dfrac{\left(\Omega_{\text{M},0}-1\right)}{1-\left(1+p\right)e^{-p}}  p^2 \exp \left[-p \sqrt{\dfrac{T}{T_0}}\right]}.
\end{equation}
Using the field equations, the expression can be re-written as
\begin{widetext}

\begin{equation}
w_{\text{exo}} = -\left[1+\dfrac{\Omega_{\text{M},0} a^{-3} p^2 \exp \left[-p \sqrt{\dfrac{T}{T_0}}\right]}{2\left\lbrace 1-\left(1+p\sqrt{\dfrac{T}{T_0}}\right)\exp \left[-p \sqrt{\dfrac{T}{T_0}}\right]\right\rbrace-p^2 \dfrac{T}{T_0} \exp \left[-p \sqrt{\dfrac{T}{T_0}}\right]}\right]^{-1}.
\end{equation}

\end{widetext}
For $p > 0$, the EoS parameter is always negative; however, nothing can be inferred for $p < 0$ since a numerical solution must be carried out to determine the nature of the EoS parameter. At current times, the value of the EoS parameter is given by
\resizebox{0.97\linewidth}{!}{
  \begin{minipage}{\linewidth}
\begin{equation}
w_{\text{exo}}\left(t_0\right) = -\left\lbrace 1+\dfrac{\Omega_{\text{M},0} p^2 e^{-p}}{2\left[ 1-\left(1+p\right)e^{-p}\right]-p^2 e^{-p}}\right\rbrace^{-1}.
\end{equation}
  \end{minipage}
}
As discussed in the previous section, since current data models do not use Lagrangian functionals used here but seem to indicate that $w_{\text{exo}} \approx -1$, we consider those models which at current times have a EoS parameter close to such value. We first however obtain a true constraint from the fact that universe is currently accelerating, i.e. from the deceleration parameter. For this model, the deceleration parameter is
\begin{widetext}

\begin{equation}
q = -1+3\frac{1+\dfrac{(\Omega_{\text{M},0}-1)}{1-(1+p)e^{-p}} \dfrac{T_0}{T} \left\lbrace 1-\left(1+ p \sqrt{\dfrac{T}{T_0}}\right)\exp \left[-p \sqrt{\dfrac{T}{T_0}}\right]\right\rbrace}{2+\dfrac{(\Omega_{\text{M},0}-1)}{1-(1+p)e^{-p}}  p^2 \exp \left[-p \sqrt{\dfrac{T}{T_0}}\right]}
\end{equation}

\end{widetext}
To set constraints on the parameter $p$, one requires that at current times $q\left(t_0\right) < 0$. Within this model, the deceleration parameter turns out to be given by
\begin{equation}
q\left(t_0\right) = -1+3\frac{\Omega_{\text{M},0}}{2+\dfrac{(\Omega_{\text{M},0}-1)}{1-(1+p)e^{-p}}  p^2e^{-p}}.
\end{equation}
By again assuming that $\Omega_{\text{M},0} \approx 0.3$ leads to $p < -1.18$ or $p > 0.68$. Since $p > 0$ have a negative EoS parameter for the exotic fluid, we opt for such models and consider two in particular, being $p = 2$ ($w_{\text{exo}} \approx -0.80$) and $p = 5$ ($w_{\text{exo}} \approx -0.97$). 

In the following figures, we analyse some of the features for power-law cosmologies with $\Lambda$CDM, $p = 2$ and $p = 5$. The plots were carried out using $\Omega_{\text{M},0} = 0.3$ and $H_0 = \left(\SI{14.4}{\giga\year}\right)^{-1}$. The solid curve represents $\Lambda$CDM, whilst the dotted and dashed curves represent $p = 2$ and $p = 5$ respectively. 

\begin{figure}[h!]
\includegraphics[width=0.49\textwidth{}]{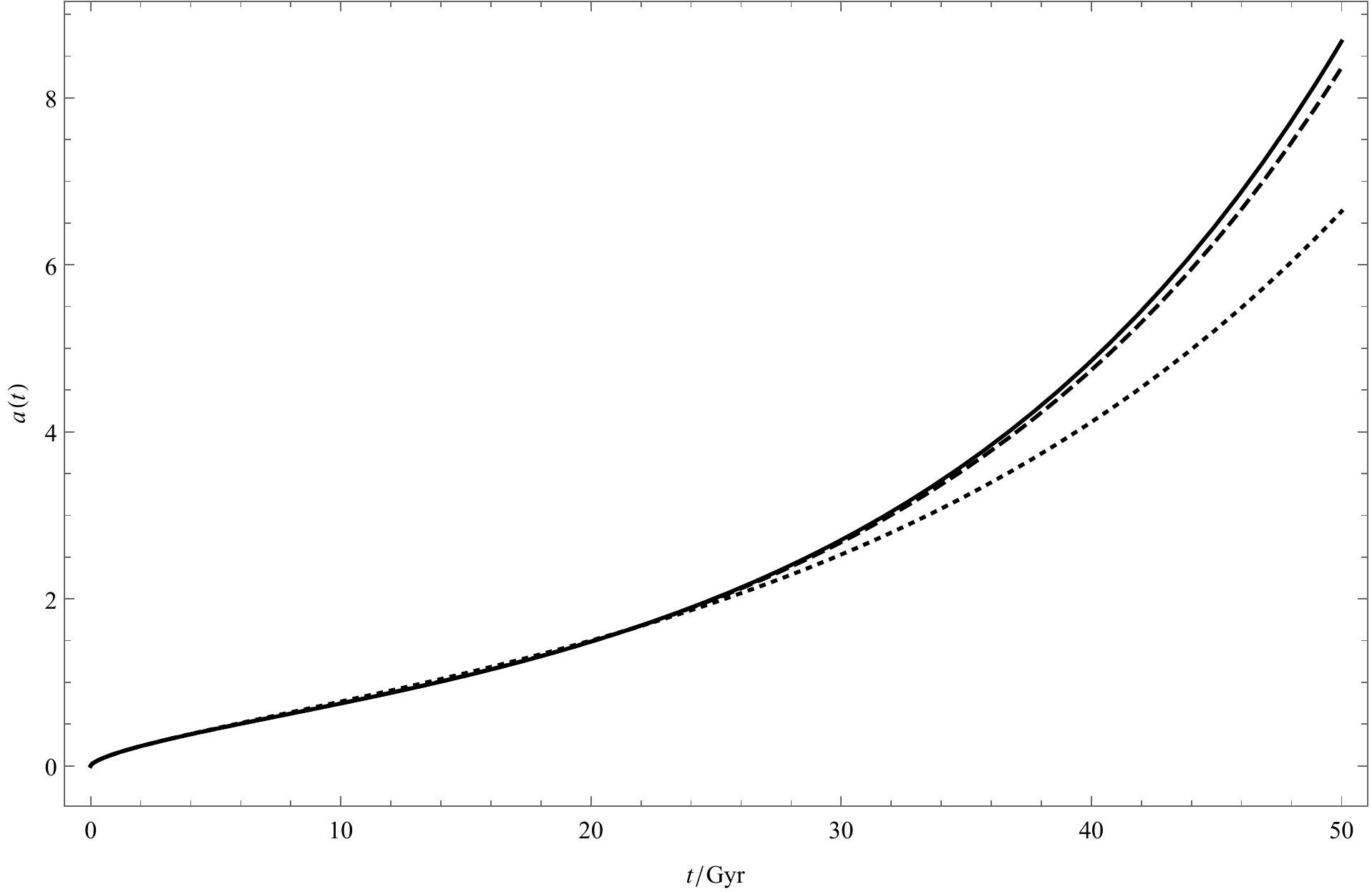}
\caption{The evolution of the scale factor $a(t)$ with cosmic time for the model $f(T) =\alpha T_0 \left(1 - \exp \left[-p \sqrt{\dfrac{T}{T_0}}\right]\right)$ for $p = 2$ and $5$ and the $\Lambda$CDM model. At early and current times, the models exhibit similar behaviour until it reaches late times where the $p = 2$ (dotted) model deviates from the $\Lambda$CDM (solid) model. The $p = 5$ (dashed) model closely mimics the $\Lambda$CDM model.}
\label{fig:fig_scalefactor_exp}
\end{figure}

We again start by examining the evolution of the scale factor $a(t)$ for each model which is given in Fig. \ref{fig:fig_scalefactor_exp}. The models at early times are almost identical but start to deviate at late times, with the most notable deviation for the $p = 2$ model.

\begin{figure}[h!]
\includegraphics[width=0.49\textwidth{}]{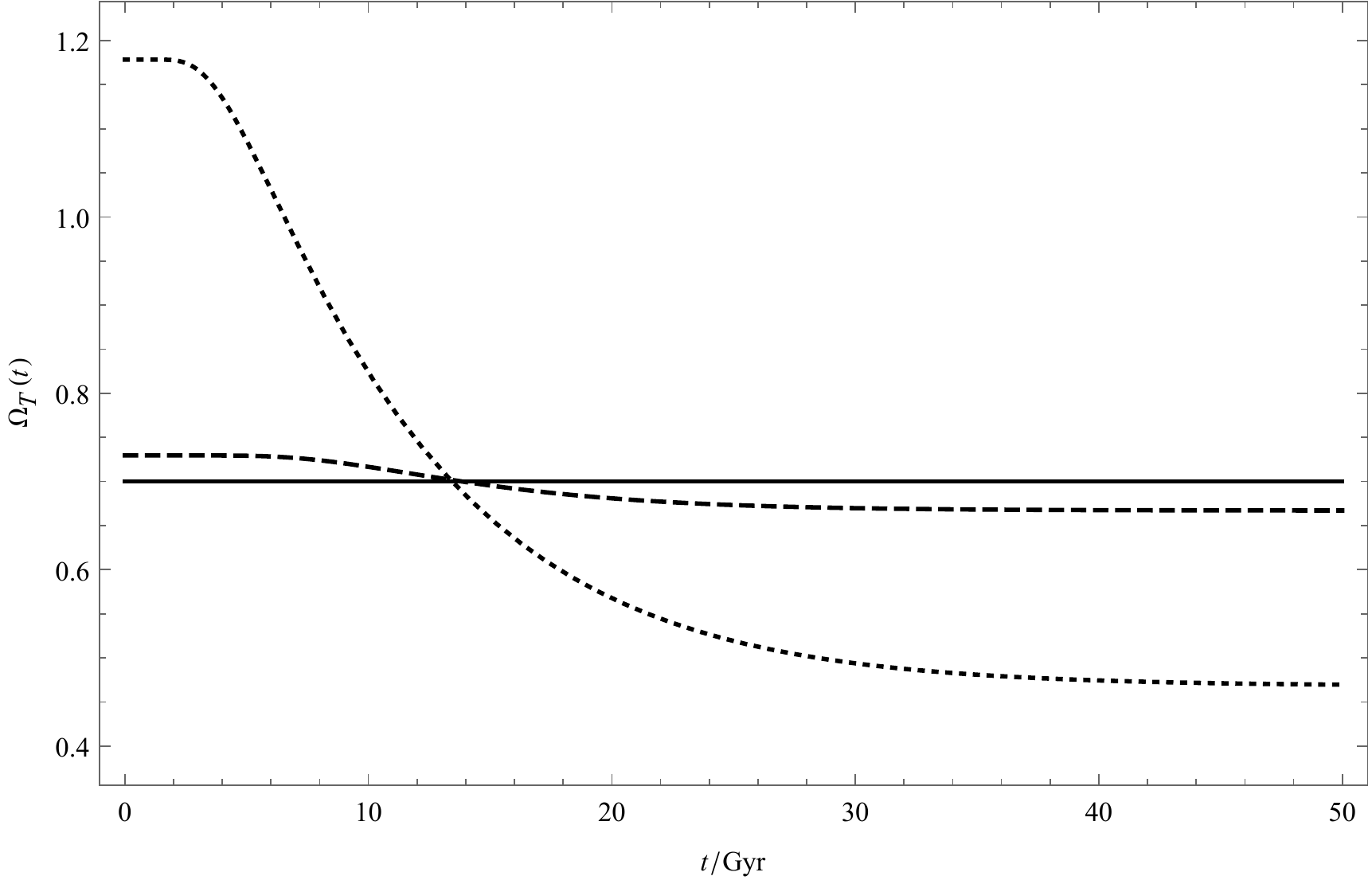}
\caption{The evolution of the torsion density parameter $\Omega_T(t)$ with cosmic time for the model $f(T) =\alpha T_0 \left(1 - \exp \left[-p \sqrt{\dfrac{T}{T_0}}\right]\right)$ for $p = 2$ and $5$ and the $\Lambda$CDM model. For these types of models, the exotic fluid's density parameter starts from a non-zero value [the $p = 2$ (dotted) model gives $\Omega_T(0) \approx 1.175$ whilst for $p = 5$ (dashed) model gives $\Omega_T(0) \approx 0.725$] in a similar way as $\Lambda$CDM (solid) does (a constant 0.700), reaches the current time value of 0.700 and then decrease until it reaches a constant value during late times [$\Omega_T\left(t = t_\infty\right) \approx 0.475$ for the $p = 2$ model and $\Omega_T\left(t = t_\infty\right) \approx 0.675$ for the $p = 5$ model]. The $p = 2$ model is marginally different from the $p = 5$ and $\Lambda$CDM models, whilst the $p = 5$ model closely mimics the $\Lambda$CDM behaviour.}
\label{fig:fig_OmegaT_exp}
\end{figure}

The variation of $\Omega_T$ with cosmic time is shown in Fig. \ref{fig:fig_OmegaT_exp}. In contrast with the power-law model, the exotic fluid's density parameter initially is non-zero here, similar to the cosmological constant [for the $p = 2$ model, $\Omega_T(0) \approx 1.175$ whilst for the $p = 5$ model, $\Omega_T(0) \approx 0.725$]. This value decreases until it reaches the current observed value of 0.7 and decreases until it reaches a constant value at late times [for the $p = 2$ model, $\Omega_T\left(t = t_\infty\right) \approx 0.475$ whilst for the $p = 5$ model, $\Omega_T\left(t = t_\infty\right) \approx 0.675$].

\begin{figure}[h!]
\includegraphics[width=0.49\textwidth{}]{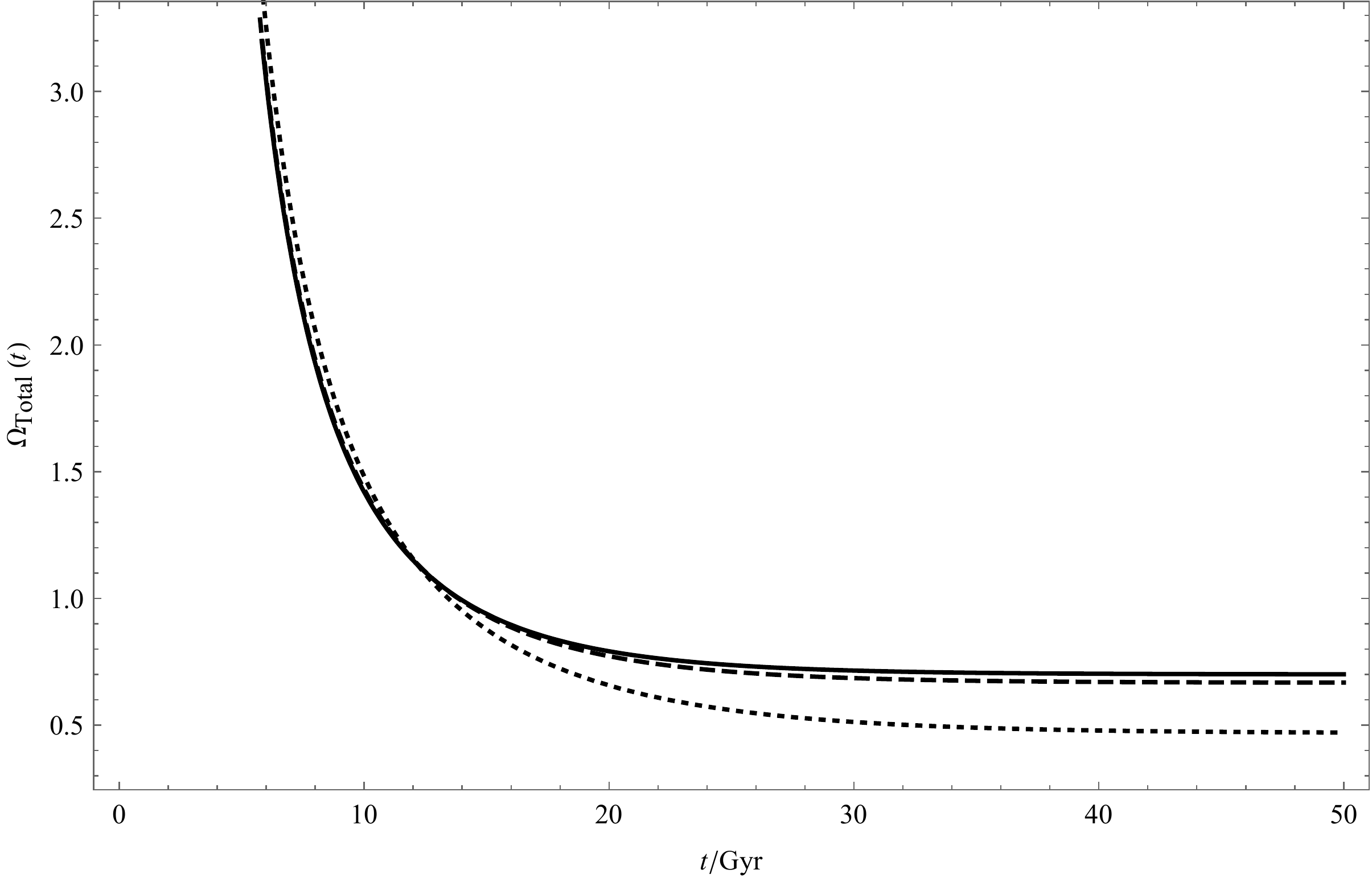}
\caption{The evolution of the total density parameter $\Omega_{\text{Total}}(t)$ with cosmic time for the model $f(T) =\alpha T_0 \left(1 - \exp \left[-p \sqrt{\dfrac{T}{T_0}}\right]\right)$ for $p = 2$ and $5$ and the $\Lambda$CDM model. Each model mimics the behaviour of $\Lambda$CDM (solid); it starts at large values due to the matter density being dominant over the exotic fluid's density, then it decreases and reaches 1 at present time and keeps on decreasing until it reaches a constant value at future times (since the matter density becomes irrelevant and the exotic fluid's density becomes constant). Such future time value is noticeably different for the $p = 2$ (dotted) model compared with the $p = 5$ and $\Lambda$CDM models $\left[\Omega_{\text{Total}}\left(t = t_\infty\right) \approx 0.50\right]$. On the other hand, the $p = 5$ (dashed) model greatly mimics the $\Lambda$CDM model [$\Omega_{\text{Total}}\left(t = t_\infty\right) \approx 0.65$ for $p = 5$ and $\Omega_{\text{Total}}\left(t = t_\infty\right) \approx 0.70$ for $\Lambda$CDM].}
\label{fig:fig_OmegaTotal_exp}
\end{figure}

Found the evolution of $\Omega_T$, we now look at the evolution of the total density parameter $\Omega_{\text{Total}} \equiv \Omega_\text{M} + \Omega_T$, which is shown in Fig. \ref{fig:fig_OmegaTotal_exp}. The two exponential models both mimic the $\Lambda$CDM model to a degree; initially, they have a high value since the matter density is much more dominant than the exotic fluid's density,  then the total density approaches a constant value because the exotic fluid's density becomes constant and the matter density becomes irrelevant (much smaller than the exotic fluid's density). What differs from the models is the limiting values, which are as follows: the $p = 2$ model gives $\Omega_{\text{Total}}\left(t = t_\infty\right) \approx 0.50$ whilst for the $p = 5$ model, $\Omega_{\text{Total}}\left(t = t_\infty\right) \approx 0.65$; the $\Lambda$CDM model gives $\Omega_{\text{Total}}\left(t = t_\infty\right) \approx 0.70$.

\begin{figure}[h!]
\includegraphics[width=0.49\textwidth{}]{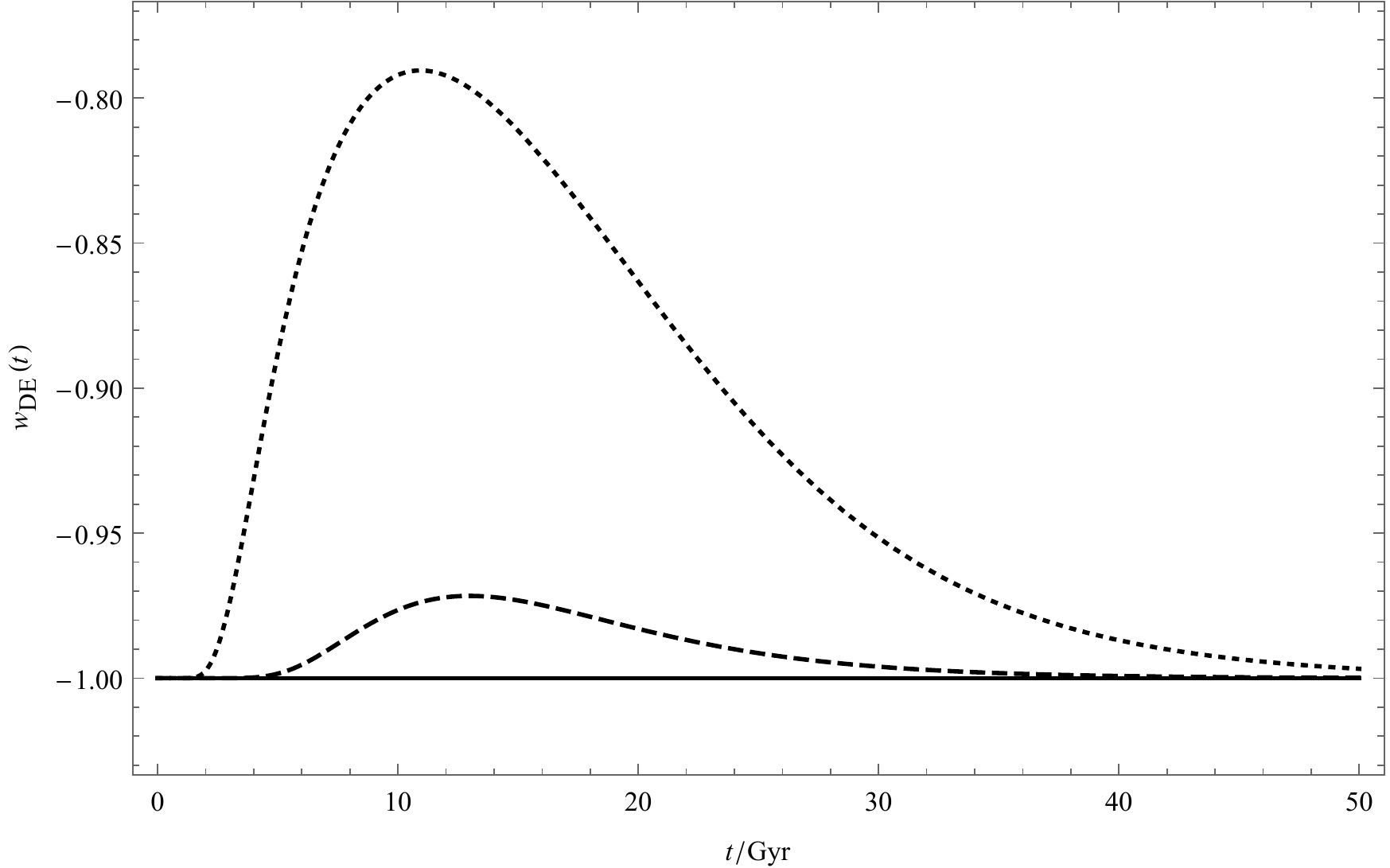}
\caption{The evolution of the exotic fluid's EoS parameter $w_{\text{exo}}(t)$ with cosmic time for the model $f(T) =\alpha T_0 \left(1 - \exp \left[-p \sqrt{\dfrac{T}{T_0}}\right]\right)$ for $p = 2$ and $5$ and the $\Lambda$CDM model. The $p = 2$ (dotted) and $p = 5$ (dashed) models greatly differ from the behaviour of $\Lambda$CDM (solid) during current times, in which the fluid behaves as a non-phantom fluid. Both models however begin and end with an EoS parameter value of $-1$ indicating that at such instances, the fluid behaves as a cosmological constant as $\Lambda$CDM.}
\label{fig:fig_wDE_exp}
\end{figure}

Fig. \ref{fig:fig_wDE_exp} shows the variation of $w_{\text{exo}}$ with cosmic time. One notes that initially and during late times, the models start and end with $w_{\text{exo}} = -1$ which mimics the $\Lambda$CDM value. However between these times, the EoS parameter describes the exotic fluid to behave as a non-phantom fluid with a maximum peak just before current time $t_0 \approx \SI{14}{\giga\year}$ (for the $p = 2$ model, the maximum occurs at $t \approx \SI{11}{\giga\year}$ having value $w^{\text{max}}_\text{exo} \approx -0.79$ whilst for the $p = 5$ model, this is achieved at $t \approx \SI{13}{\giga\year}$ having value $w^{\text{max}}_\text{exo} \approx -0.97$).

\begin{figure}[h!]
\includegraphics[width=0.49\textwidth{}]{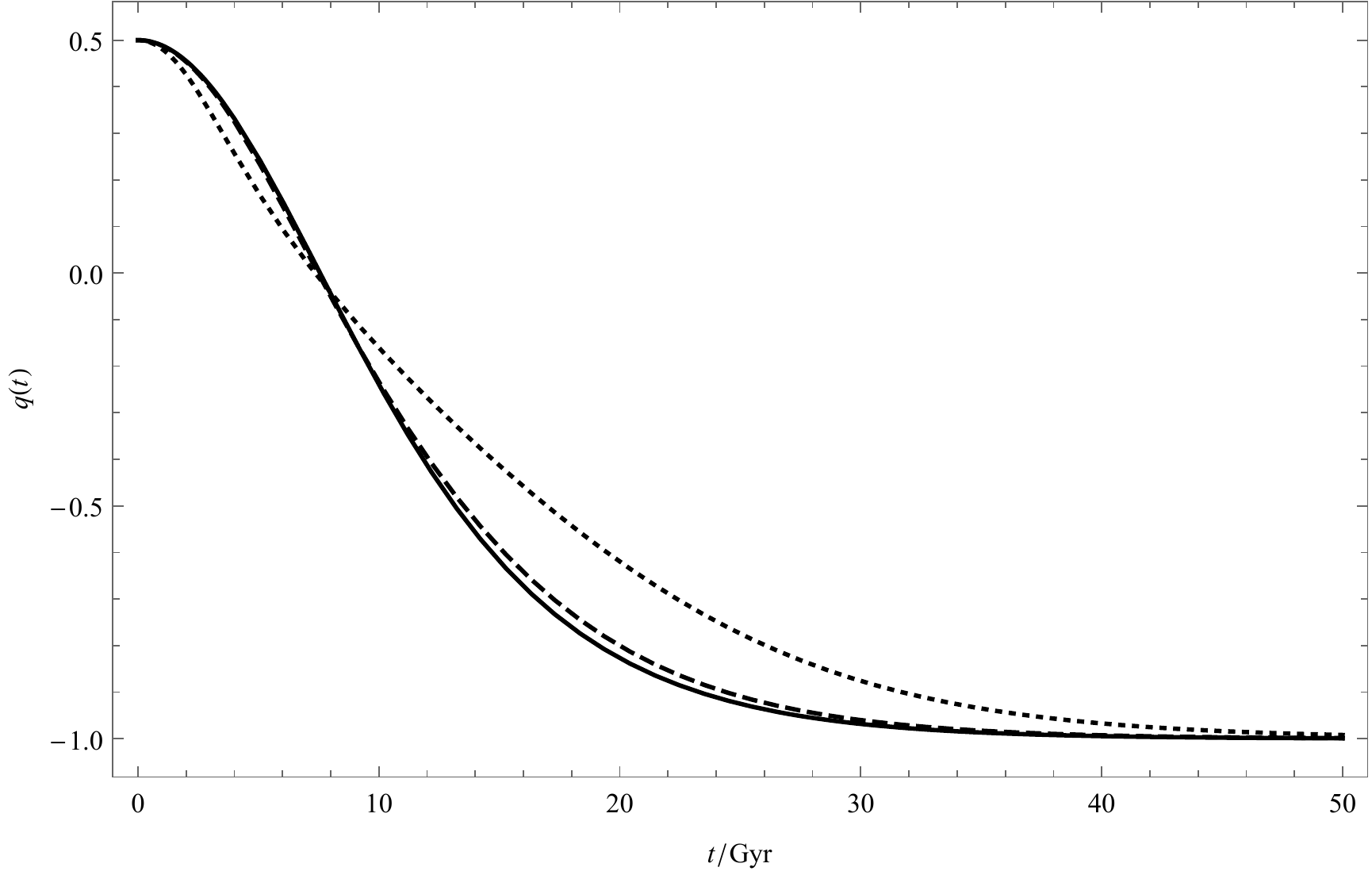}
\caption{The evolution of the deceleration parameter $q(t)$ with cosmic time for the model $f(T) =\alpha T_0 \left(1 - \exp \left[-p \sqrt{\dfrac{T}{T_0}}\right]\right)$ for $p = 2$ and $5$ and the $\Lambda$CDM model. All models exhibit the same starting value of 0.5 and late times limiting value of $-1.0$. Each model transitions to an accelerating universe at approximately the same time $(t \approx \SI{8}{\giga\year})$. The $p = 5$ (dashed) model greatly mimics the $\Lambda$CDM (solid) evolution whilst the $p = 2$ (dotted) model, although having a similar behaviour, differs from the latter.}
\label{fig:fig_dec_exp}
\end{figure}

The evolution of the deceleration parameter is shown in Fig. \ref{fig:fig_dec_exp}. One notes that the models have similar initial and late time behaviour, i.e. $q(0) = 0.5$ and $q\left(t = t_\infty\right) = -1.0$ being the same as $\Lambda$CDM. Furthermore, the overall behaviour is also similar to $\Lambda$CDM, with the major difference being that the various values of $q(t)$ occur at different times. Moreover, the transition between non-accelerating and accelerating happens at approximately the same time $(t \approx \SI{8}{\giga\year})$. 

\begin{figure}[h!]
\includegraphics[width=0.49\textwidth{}]{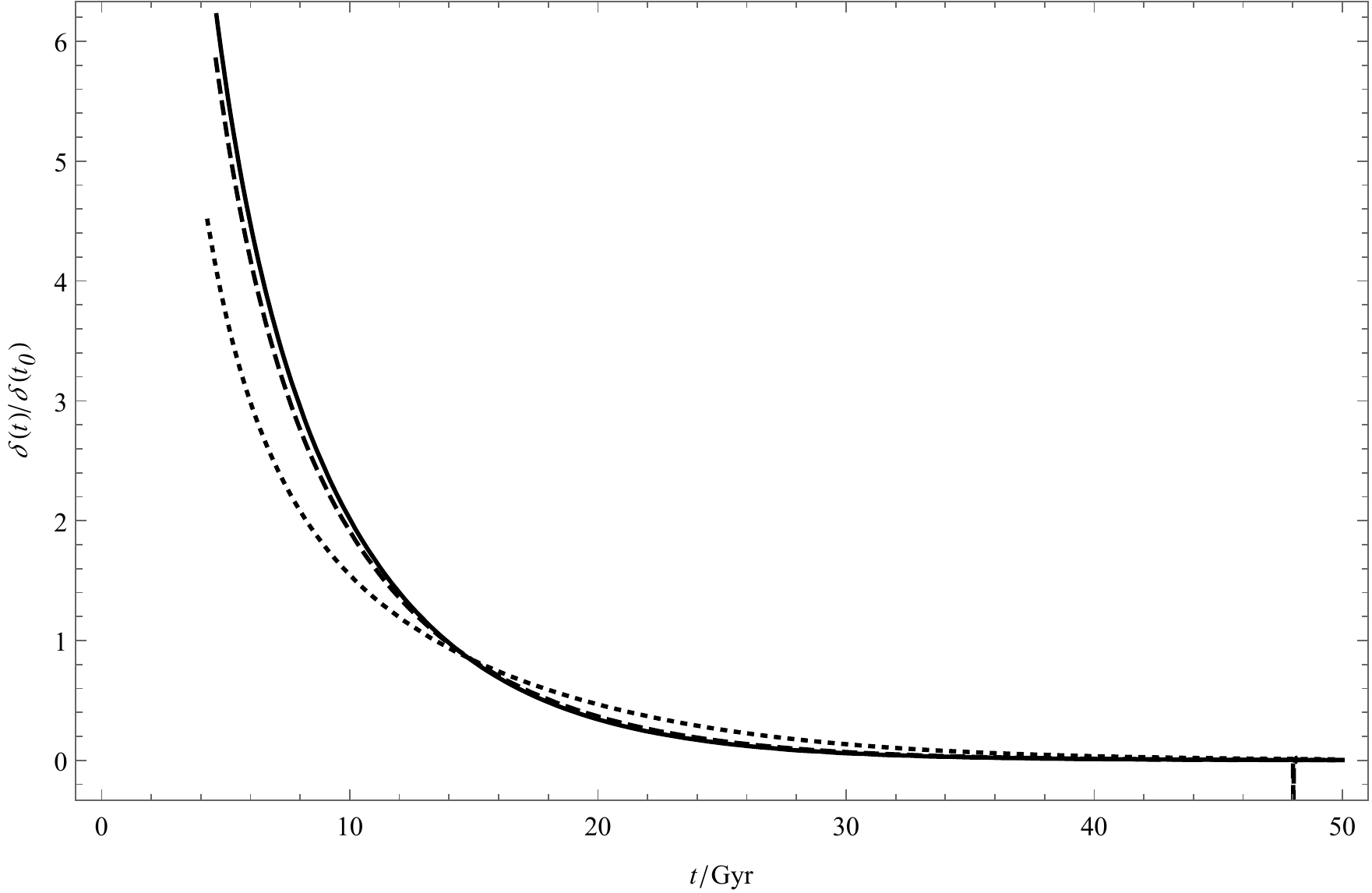}
\caption{The evolution of the ratio of the Hubble perturbation parameter $\delta(t)$ to its current value $\delta\left(t_0\right)$ with cosmic time for the model $f(T) =\alpha T_0 \left(1 - \exp \left[-p \sqrt{\dfrac{T}{T_0}}\right]\right)$ for $p = 2$ and $5$ and the $\Lambda$CDM model. In this case, the models exhibit the same behaviour as $\Lambda$CDM (solid), starting from large values, decreasing and approaching approximately 1 at around present time and then decay to zero at late times, meaning these perturbations are stable and do not persist. The $p = 5$ (dashed) model is very close to the $\Lambda$CDM model behaviour whilst the $p = 2$ (dotted) model differs from the other two models at earlier times.}
\label{fig:fig_delta_exp}
\end{figure}

\begin{figure}[h!]
\includegraphics[width=0.49\textwidth{}]{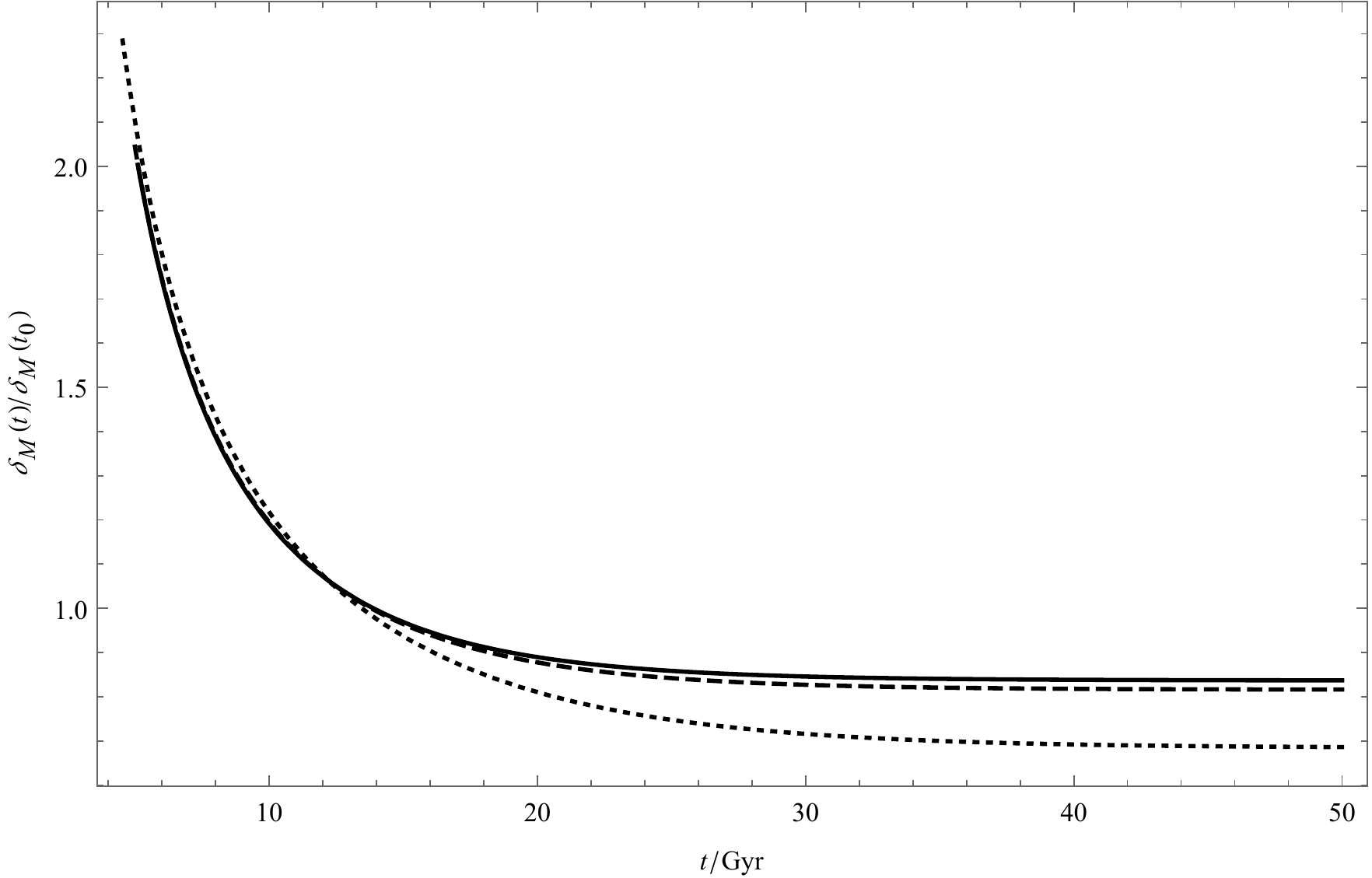}
\caption{The evolution of the ratio of the matter perturbation parameter $\delta_\text{M}(t)$ to its current value $\delta_\text{M} \left(t_0\right)$ with cosmic time for the model $f(T) =\alpha T_0 \left(1 - \exp \left[-p \sqrt{\dfrac{T}{T_0}}\right]\right)$ for $p = 2$ and $5$ and the $\Lambda$CDM model. Both models mimic the $\Lambda$CDM (solid) behaviour; each start at a large value at early times, decay to approximately 1 at around today's time and then keeps on decreasing until reaching a constant value, meaning that the perturbation is stable but persists. The difference lies in the values, but with the major difference occurring for $p = 2$ (dotted) [for $\Lambda$CDM, $\delta(t)/\delta\left(t_0\right) \approx 0.85$ whilst for $p = 2$, $\delta(t)/\delta\left(t_0\right) \approx 0.70$ and for $p = 5$ (dashed), $\delta(t)/\delta\left(t_0\right) \approx 0.80$].}
\label{fig:fig_deltaM_exp}
\end{figure}

Lastly, the evolution of the perturbations of the Hubble and density are shown in Figs. \ref{fig:fig_delta_exp} and \ref{fig:fig_deltaM_exp} respectively. Starting with the Hubble perturbations, such perturbations decay with time and completely vanish at late times similar to $\Lambda$CDM. On the other hand, the matter perturbations initially decay with time until reaching a limiting value at late times which is again similar to $\Lambda$CDM [for $\Lambda$CDM, $\delta(t)/\delta\left(t_0\right) \approx 0.85$ whilst for $p = 2$, $\delta(t)/\delta\left(t_0\right) \approx 0.70$ and for $p = 5$, $\delta(t)/\delta\left(t_0\right) \approx 0.80$].

\section{VI. Conclusion}

In this paper, we investigated the stability of $f(T)$ gravity. The perturbed field equations have been derived and solved analytically to give Eqs. \eqref{eq:deltam-solution} and \eqref{eq:delta-solution}. Both solutions are dependent on the Hubble parameter, which ultimately determines whether such solutions are stable. This was analysed under power-law and de-Sitter evolutions, as well as under two separate $f(T)$ models.

For power-law and de-Sitter evolutions, the perturbations are stable. For the Hubble perturbation, it is either decaying to zero at late times (power-law) or being identically zero (de-Sitter). On the other hand, the matter perturbation either decays to zero at late times (power-law) or is constant (de-Sitter) meaning that such perturbations persist. Although such solutions are stable, a problem arises due to $tt$-Friedmann equation Eq. \eqref{eq:00-zero} since this restricts the possible functions of $f(T)$ where power-law and de-Sitter evolution can be considered. In fact, we have shown that the only functions which allow these types of evolutions are TEGR and rescaled TEGR. Nonetheless, since the universe is composed of a mix of fluids and hence the history of the universe is not described by either a power-law or exponential at all times (these serve as a good approximation at certain eras), we instead have set our focus on non-trivial $f(T)$ functions which mimic $\Lambda$CDM, which is a much more accurate description of the history of the universe.

The first model considered in this paper is the power-law model. By examining the exotic fluid's EoS parameter and deceleration parameter at current times, it was concluded that two possible non-trivial models can be obtained, $n = -1$ and $n = -2$. Such models were found to be stable, and mimic the behaviour of $\Lambda$CDM. The Hubble perturbation decays with cosmic time and approaches zero at late times in the same as $\Lambda$CDM does. On the other hand, the matter perturbation also decays with time but tends to a limiting value at late times, again as $\Lambda$CDM does, meaning that such perturbations persist at late times. However, both $n = -1$ and $n = -2$ models differ significantly from the limiting $\Lambda$CDM value [for $\Lambda$CDM, $\delta(t)/\delta\left(t_0\right) \approx 0.850$ whilst for $n = -1$ and $n = -2$, $\delta(t)/\delta\left(t_0\right) \approx 0.925$].

The exponential model is the second model considered in this paper. In this case, two possible models which agree with data are considered, $p = 2$ and $p = 5$. The models are found to be stable and exhibit the same behaviour as $\Lambda$CDM. For the Hubble perturbation, each model decays to zero at late times whilst matter perturbations decay to a constant value at late times, and hence such perturbations persist. The $p = 2$ model deviates the most compared to $\Lambda$CDM for each parameter considered whilst the $p = 5$ was the closest to mimic the latters behaviour.

All in all, the $p = 5$ exponential model provided the closest behaviour to $\Lambda$CDM when all models are considered comparatively. Nonetheless, other $f(T)$ functions which may provide a better explanation and agree to the observables of the universe which are not considered in this paper exist. As long as such functions are able to give stable solutions, it helps constrain the theory's parameters. At the same time, it helps in restricting the number of possible functions which $f(T)$ theory permits, ultimately contributing in defining a proper Lagrangian for describing gravity.

\section*{Acknowledgements}

The research work disclosed in this paper is partially funded by the ENDEAVOUR Scholarships Scheme.

\end{document}